# Internal flows and energy circulation in light beams

**Aleksandr Bekshaev[1], Konstantin Y. Bliokh[2] and Marat Soskin[3]**

[1] I.I. Mechnikov National University, Dvorianska 2, Odessa, 65082, Ukraine;
E-mail: bekshaev@onu.edu.ua

[2] Applied Optics Group, School of Physics, National University of Ireland, Galway, Galway, Ireland

[3] Institute of Physics, National Academy of Science of Ukraine, Prospect Nauki 46, Kiev, 03028 Ukraine

**Abstract**

We review optical phenomena associated with the internal energy redistribution which accompany propagation and transformations of monochromatic light fields in homogeneous media. The total energy flow (linear-momentum density, Poynting vector) can be divided into spin part associated with the polarization and orbital part associated with the spatial inhomogeneity. We give general description of the internal flows in the coordinate and momentum (angular spectrum) representations for both nonparaxial and paraxial fields. This enables one to determine local densities and integral values of the spin and orbital angular momenta of the field. We analyse patterns of the internal flows in standard beam models (Gaussian, Laguerre-Gaussian, flat-top beam, etc.), which provide an insightful picture of the energy transport. The emphasize is made to the singular points of the flow fields. We describe the spin-orbit and orbit-orbit interactions in the processes of beam focusing and symmetry breakdown. Finally, we consider how the energy flows manifest themselves in the mechanical action on probing particles and in the transformations of a propagating beam subjected to a transverse perturbation.

**Abbreviations**

| | |
|---|---|
| AM | Angular momentum |
| LG | Laguerre-Gaussian |
| OAM | Orbital angular momentum |
| OFD | Orbital flow density |
| OOI | Orbit-orbit interaction |
| SAM | Spin angular momentum |
| SFD | Spin flow density |
| SOI | Spin-orbit interaction |

## 1. Introduction.

Standard approaches to characterization of light beams deal with their exterior: a beam field is described as it 'looks' outside – for an 'external' observer. Usual beam parameters either characterize a beam 'in a whole' (power, momentum, beam size and divergence angle) or describe its 'shape' via certain spatial distributions (amplitude, phase, polarization state, etc.). For a long time, such a pictorial representation looked adequate from the fundamental point of view and was quite sufficient for applications. The situation began to change after the beams with angular momentum (AM) have become an object of rapt attention (see, e.g., reviews [1–5]). This new and, at first glance, extraordinary mechanical property of light (despite that it occupied its fitting place among the fundamental physical concepts since the times of Poynting [6] and Beth [7]) stimulated especial interest to the physical foundations of the 'immediately observable' characteristics of an optical field. In the course of progressively developing investigations, it became clear that the usual beam parameters provide only rough and, sometimes, distorted picture of internal processes that constitute a real 'inner life' of a light beam. These processes are related to the fundamental dynamical and geometrical aspects of light fields, and are associated with the permanent energy redistribution inside the beam 'body', which underlies the beam evolution and transformations. The internal energy flows provide a natural and efficient way for 'peering' into the light fields and studying their most intimate and deep features.

It would be misleading, however, to think that internal flows attracted no attention before the 'AM era'. Episodically, they appeared in the focus of interest, and each time it was associated with some examples of singular behaviour of light fields, which is not surprising as in such cases the lack of adequate instruments within the traditional descriptive arsenal was the most perceptible. As early as in 1919, Ignatowskii [8], in his theoretical study of the near-focus field pattern, revealed the possibility of backward energy flow. Later, the vortex structures of the Poynting vector fields near the Airy rings in the focal region were subject of detailed analysis [9,10]. Other early recognized examples of the vortex flow in electromagnetic fields appear due to interference between the incident and reflected waves in the course of the plane-wave diffraction on a half-plane perfectly conducting screen [11] and in the process of total reflection [12]. These instances demonstrate the suitability of internal flows for analysis of singular optical phenomena, and it is not striking that the fast development of 'singular optics' [2,5,13] in the past years supplies another powerful stimulus for the energy flow investigations.

Besides these fundamental motives, the energy flows offer practical advantages in application to many other problems of modern optics. Traditional methods of the description of light fields are predominantly based on the geometric-optics or paraxial approaches, and, therefore, they become unsuitable for a number of modern topics: near-field optics, strongly focused beams, micro- and nanooptics. In this context, the energy flows provide a natural system of sensitive parameters which are not restricted by any approximations and can be employed for analysis of fine internal features of an arbitrary light field. Importantly, the energy flows represent immediately observable quantities with explicit and unambiguous physical meaning and they enable one to separate the spatial (orbital) and polarization (spin) degrees of freedom of light in the general nonparaxial case.

The growing interest to the internal flows has inspired numerous works treating various aspects of this topical and appealing matter. In this review we aim to summarize and systematically represent previous results from the unified position, addressing the mathematical description, physical interpretation and special features of the energy flows in light beams. In order to concentrate at the most fundamental aspects without unnecessary complications, we restrict the consideration to the monochromatic fields propagating in the free space. The paper is organized as follows. Section 2 introduces basic theoretical description of the energy flow, its 'structural' division into the spin and orbital parts, and calculation of the corresponding AM constituents for an arbitrary nonparaxial field. An important case of paraxial beams is considered in detail in section 3. General features of the internal flow patterns are illustrated in section 4 by the examples of Gaussian, Laguerre-Gaussian and Bessel beams. An opposite case – complicated inhomogeneous fields, especially with stochastic properties – can be studied with the help of networks of the optical flow singularities ('singular skeletons') whose properties are briefly discussed in section 5. In section 6 we consider interactions and mutual conversions of the spin and orbital AMs and energy flows. Problems of immediate physical manifestations and experimental observations of the internal flows are only at the early stages of their consideration; some aspects are outlined in rather illustrative section 7, which can be interesting to readers that prefer to stand aside the detailed theory. The review is finalized by conclusions.

## 2. General definitions and basic equations

We consider classical monochromatic electromagnetic field in free space and use the Gaussian system of units. The space coordinates are given by 3D radius-vector $\mathbf{R} = \mathbf{r} + \mathbf{e}_z z$, where the longitudinal coordinate $z$ is associated with the predominant direction of the beam propagation and $\mathbf{r} = \mathbf{e}_x x + \mathbf{e}_y y$ is the transverse radius-vector; $\mathbf{e}_x$, $\mathbf{e}_y$ and $\mathbf{e}_z$ are the unit vectors of the Cartesian frame. The field is supposed to be spatially coherent, which enables us to employ the complex representation of the real electric and magnetic fields, $\boldsymbol{E}(\mathbf{R},t)$ and $\boldsymbol{H}(\mathbf{R},t)$, oscillating with frequency $\omega$:

$$\boldsymbol{E}(\mathbf{R},t) = \mathrm{Re}\left[\mathbf{E}(\mathbf{R})\exp(-i\omega t)\right], \quad \boldsymbol{H}(\mathbf{R},t) = \mathrm{Re}\left[\mathbf{H}(\mathbf{R})\exp(-i\omega t)\right]. \tag{2.1}$$

The field energy averaged over the period of oscillations is distributed with volume density [14]

$$w = \frac{g}{2}\left(|\mathbf{E}|^2 + |\mathbf{H}|^2\right), \tag{2.2}$$

whereas the measure of the energy flow (more precisely, the flow density) is given by the time-averaged Poynting vector [14]

$$\mathbf{S} = cg\,\mathrm{Re}\left[\mathbf{E}^* \times \mathbf{H}\right]. \tag{2.3}$$

Here $g$ is the constant factor which equals $g = (8\pi)^{-1}$ in the Gaussian system and $c$ is the velocity of light. The association of the Poynting vector with the energy flow of an electromagnetic field is common but somewhat ambiguous. This interpretation rests upon the Poynting theorem

$$\frac{\partial}{\partial t}\int_V w(\mathbf{R})\, d^3\mathbf{R} = -\oint_{F_V} \mathbf{S}\cdot d\mathbf{F} \tag{2.4}$$

which relates diminution of electromagnetic energy within volume $V$ with the flux of vector **S** through the boundary of this volume $F_V$ [14]. The physical contents of equality (2.4) will not change if arbitrary solenoidal (zero-divergence) field is added to **S** (2.3). A physically meaningful realization of such a possibility was suggested by Green and Wolf [15,16] who introduced modified definitions of the energy and energy flow densities which differ from the equations (2.2) and (2.3) in small (subwavelength) scale but coincide with them in the average. Nonetheless, below we adhere to the traditional definitions of the energy density (2.2) and flow (2.3); a short discussion of the physical grounds and alternatives can be found in [17,18] (see also section 7.1 below).

Note that due to requirements of the special relativity [18–20] the Poynting vector (2.3) also expresses the momentum density of the field:

$$\mathbf{p} = \frac{1}{c^2}\mathbf{S} = \frac{g}{c}\mathrm{Re}\left[\mathbf{E}^* \times \mathbf{H}\right]. \tag{2.5}$$

This links the Poynting vector to the dynamical attributes of light. In what follows, we ignore the difference between **p** and **S** and use the terms “momentum density” and “energy flow density” as synonyms. In accordance with (2.5), the density of the AM of light is given by

$$\mathbf{j} = \mathbf{R}\times\mathbf{p} = \frac{g}{c}\mathrm{Re}\left[\mathbf{R}\times\left(\mathbf{E}^* \times \mathbf{H}\right)\right]. \tag{2.6}$$

For some problems it is more convenient to define the AM with respect to the $z$-axis; in this case the 3D radius-vector in (2.6) is replaced by its transverse projection: $\mathbf{R} \to \mathbf{r}$.

By using the Maxwell equations for monochromatic fields [14], the momentum density (2.5) can be written as

$$\mathbf{p} = \frac{g}{\omega}\mathrm{Im}\left[\mathbf{E}^* \times \left(\nabla\times\mathbf{E}\right)\right] = \frac{g}{\omega}\mathrm{Im}\left[\mathbf{H}^* \times \left(\nabla\times\mathbf{H}\right)\right] = \mathbf{p}_S + \mathbf{p}_O \tag{2.7}$$

where

$$\mathbf{p}_S = \frac{g}{4\omega}\mathrm{Im}\left[\nabla\times\left(\mathbf{E}^*\times\mathbf{E} + \mathbf{H}^*\times\mathbf{H}\right)\right], \tag{2.8}$$

$$\mathbf{p}_O = \frac{g}{2\omega}\mathrm{Im}\left[\mathbf{E}^*\cdot(\nabla)\mathbf{E} + \mathbf{H}^*\cdot(\nabla)\mathbf{H}\right] \tag{2.9}$$

represent the spin and orbital constituents of the total energy flow. Hereinafter, these quantities are referred to as ‘spin flow density’ (SFD) and ‘orbital flow density’ (OFD), respectively. The spin-orbit decomposition (2.7) – (2.9) was proposed by Bekshaev and Soskin for paraxial beams [21] and extended by Berry to general nonparaxial fields [17]. It is based on the fact that after substitution into (2.6) the term $\mathbf{p}_S$ gives rise to the spin AM (SAM) of the field, while $\mathbf{p}_O$ produces the orbital AM (OAM) (see also [22–25]). In equation (2.9) we used notation

$$\mathbf{E}^*\cdot(\nabla)\mathbf{E} = E_x^*\nabla E_x + E_y^*\nabla E_y + E_z^*\nabla E_z, \tag{2.10}$$

and expressions (2.8) and (2.9) are presented in the form which adopts the ‘electric-magnetic democracy’ [17]. In the absence of electric charges and currents, the total energy flow, as well as its spin and orbital parts, are solenoidal:

$$\nabla\cdot\mathbf{p}=\nabla\cdot\mathbf{p}_S=\nabla\cdot\mathbf{p}_O=0, \tag{2.11}$$

i.e., the energy flow lines are always continuous. Important characteristics of these vector fields are their vorticities [17]:

$$\boldsymbol{\Omega}_S=\nabla\times\mathbf{p}_S=-\frac{g}{4\omega}\operatorname{Im}\nabla^2\left(\mathbf{E}^*\times\mathbf{E}+\mathbf{H}^*\times\mathbf{H}\right), \tag{2.12}$$

$$\boldsymbol{\Omega}_O=\nabla\times\mathbf{p}_O=\frac{g}{2\omega}\operatorname{Im}\left[\nabla\mathbf{E}^*\cdot(\times\nabla)\mathbf{E}+\nabla\mathbf{H}^*\cdot(\times\nabla)\mathbf{H}\right]. \tag{2.13}$$

In equation (2.13), the scalar product relates the field vectors, whereas the vector product relates operators $\nabla$, so that in Cartesian representation

$$\nabla\mathbf{E}^*\cdot(\times\nabla)\mathbf{E}=\nabla E_x^*\times\nabla E_x+\nabla E_y^*\times\nabla E_y+\nabla E_z^*\times\nabla E_z\,.$$

An alternative representation of the decomposition (2.7) – (2.9) was proposed by Berry [17], Li [23] and Bliokh *et al.* [24], which is based upon the plane-wave expansion of the fields:

$$\mathbf{E}(\mathbf{R})=\frac{1}{2\pi}\int\limits_{|\mathbf{k}|=k}\tilde{\mathbf{E}}(\mathbf{k})e^{i\mathbf{k}\cdot\mathbf{R}}d^2\mathbf{k},\quad \mathbf{H}(\mathbf{R})=\frac{1}{2\pi}\int\limits_{|\mathbf{k}|=k}\tilde{\mathbf{H}}(\mathbf{k})e^{i\mathbf{k}\cdot\mathbf{R}}d^2\mathbf{k}\,. \tag{2.14}$$

Here $k=\omega/c=|\mathbf{k}|$ is the wave number and integration is performed over the hemisphere $k_z>0$ in the $\mathbf{k}$-space (we neglect contribution of evanescent modes) [24]. Transversality conditions $\nabla\cdot\mathbf{E}=\nabla\cdot\mathbf{H}=0$ lead to requirement $\mathbf{k}\cdot\tilde{\mathbf{E}}(\mathbf{k})=\mathbf{k}\cdot\tilde{\mathbf{H}}(\mathbf{k})=0$ which can be satisfied by following the known procedure [15,16,24]. Namely, we choose an auxiliary vector $\mathbf{e}_0$ (in general $\mathbf{e}_0=\mathbf{e}_0(\mathbf{k})$ but the specific choice of constant $\mathbf{e}_0$ generates special classes of non-paraxial beams which play important role in particular problems of light beam transformations [26,27]) and define two unit vectors

$$\mathbf{e}_2=\frac{\mathbf{e}_0\times\mathbf{e}_\mathbf{k}}{\left|\mathbf{e}_0\times\mathbf{e}_\mathbf{k}\right|},\quad \mathbf{e}_1=\mathbf{e}_2\times\mathbf{e}_\mathbf{k}\,, \tag{2.15}$$

where $\mathbf{e}_\mathbf{k}=\mathbf{k}/k$. The vectors ($\mathbf{e}_1$, $\mathbf{e}_2$, $\mathbf{e}_\mathbf{k}$) form a Cartesian frame in which vectors $\tilde{\mathbf{E}}(\mathbf{k})$ and $\tilde{\mathbf{H}}(\mathbf{k})$ lie in the transverse plane ($\mathbf{e}_1$, $\mathbf{e}_2$). Next, we introduce the helicity (circular-polarization) basis

$$\mathbf{e}_+(\mathbf{k})=\frac{1}{\sqrt{2}}(\mathbf{e}_1+i\mathbf{e}_2),\quad \mathbf{e}_-(\mathbf{k})=\frac{1}{\sqrt{2}}(\mathbf{e}_1-i\mathbf{e}_2) \tag{2.16}$$

in which the plane-wave components of the field (2.14) can be represented as

$$\tilde{\mathbf{E}}(\mathbf{k})=\mathbf{C}_+(\mathbf{k})+\mathbf{C}_-(\mathbf{k}),\quad \tilde{\mathbf{H}}(\mathbf{k})=-i\mathbf{C}_+(\mathbf{k})+i\mathbf{C}_-(\mathbf{k})\,, \tag{2.17}$$

$$\mathbf{C}_\sigma(\mathbf{k})=C_\sigma(\mathbf{k})\mathbf{e}_\sigma(\mathbf{k}), \tag{2.18}$$

where $\sigma=\pm1$ and $C_\sigma(\mathbf{k})$ are the scalar amplitudes of the circularly-polarized components.

Substituting (2.17) and (2.18) into (2.2) and (2.7) – (2.9), we arrive after some calculations at [24]

$$w=\frac{g}{(2\pi)^2}\sum_\sigma\int\limits_{|\mathbf{k}|=k}d^2\mathbf{k}\int\limits_{|\mathbf{k}'|=k}d^2\mathbf{k}'e^{i(\mathbf{k}-\mathbf{k}')\mathbf{R}}\left(\mathbf{C}_\sigma(\mathbf{k})\cdot\mathbf{C}_\sigma^*(\mathbf{k}')\right), \tag{2.19}$$

$$\mathbf{p}_S=\frac{g}{2\omega(2\pi)^2}\sum_\sigma\int\limits_{|\mathbf{k}|=k}d^2\mathbf{k}\int\limits_{|\mathbf{k}'|=k}d^2\mathbf{k}'e^{i(\mathbf{k}-\mathbf{k}')\mathbf{R}}\left[\mathbf{C}_\sigma(\mathbf{k})\times\mathbf{C}_\sigma^*(\mathbf{k}')\right]\times(\mathbf{k}-\mathbf{k}'), \tag{2.20}$$

$$\mathbf{p}_O=\frac{g}{2\omega(2\pi)^2}\sum_\sigma\int\limits_{|\mathbf{k}|=k}d^2\mathbf{k}\int\limits_{|\mathbf{k}'|=k}d^2\mathbf{k}'e^{i(\mathbf{k}-\mathbf{k}')\mathbf{R}}\left(\mathbf{C}_\sigma(\mathbf{k})\cdot\mathbf{C}_\sigma^*(\mathbf{k}')\right)(\mathbf{k}+\mathbf{k}'), \tag{2.21}$$

where properties of the helicity basis (2.15) and (2.16) were taken into account. Noteworthily, despite that $w$ and $\mathbf{p}$ "depend quadratically on the field, they separate into the sums of the two helicities without cross terms mixing + and – components" [17]. In fact, the interference terms containing $C_+(\mathbf{k})C_-^*(\mathbf{k}')$ vanish due to the time-averaging [15] – a special feature of the helicity basis (2.16) that does not take place in other bases, e.g., in the Cartesian frame (2.15). The plane-wave (angular-spectrum, **k-**) representation is suitable for the integral characteristics of the field [23,24]. In particular, the linear densities (per unit *z*-length) of the energy and momentum are given by

$$W = \int w(\mathbf{R})\, d^2\mathbf{r} = g\sum_\sigma \int_{|\mathbf{k}|=k} \left|\mathbf{C}_\sigma(\mathbf{k})\right|^2 \frac{d^2\mathbf{k}}{(2\pi)^2}\,, \tag{2.22}$$

$$\mathbf{P} = \int \mathbf{p}(\mathbf{R})\, d^2\mathbf{r} = \int \mathbf{p}_O(\mathbf{R})\, d^2\mathbf{r} = \frac{g}{\omega}\sum_\sigma \int_{|\mathbf{k}|=k} \mathbf{k}\left|C_\sigma(\mathbf{k})\right|^2 d^2\mathbf{k}\,, \tag{2.23}$$

where $\int(...)d^2\mathbf{r}$ means the integral over the transverse (*x*, *y*)-plane. The spin contribution to the total beam momentum is zero [23], which also follows from representation (2.8), assuming the electric and magnetic fields vanishing at the infinity. Substituting equations (2.20) and (2.21) into (2.6) and (2.7) and using delta-function properties of Fourier integrals together with the property of the helicity basis $\mathbf{e}_\sigma^* \times \mathbf{e}_\sigma = i\sigma\mathbf{e}_\mathbf{k}$, we derive the AM of the field (per unit *z*-length) [24]:

$$\mathbf{J} = \int \mathbf{R}\times(\mathbf{p}_S + \mathbf{p}_O)\, d^2\mathbf{r} = \mathbf{J}_S + \mathbf{J}_O \tag{2.24}$$

where the SAM and OAM components are

$$\mathbf{J}_S = \frac{g}{\omega}\sum_\sigma \int_{|\mathbf{k}|=k} \sigma\left|C_\sigma(\mathbf{k})\right|^2 \mathbf{e}_\mathbf{k}\, d^2\mathbf{k} = \frac{g}{\omega}\int_{|\mathbf{k}|=k} \left(\left|C_+(\mathbf{k})\right|^2 - \left|C_-(\mathbf{k})\right|^2\right)\mathbf{e}_\mathbf{k}\, d^2\mathbf{k}\,, \tag{2.25}$$

$$\mathbf{J}_O = \frac{g}{\omega}\sum_\sigma \int_{|\mathbf{k}|=k} \mathbf{C}_\sigma^*(\mathbf{k})\cdot\left(-i\mathbf{k}\times\frac{\partial}{\partial\mathbf{k}}\right)\mathbf{C}_\sigma(\mathbf{k})\, d^2\mathbf{k}\,. \tag{2.26}$$

Here, like in (2.10), the dot product relates vectors $\mathbf{C}_\sigma^*(\mathbf{k})$ and $\mathbf{C}_\sigma(\mathbf{k})$.

The SAM (2.25) does not depend on the reference point and is completely intrinsic, while the OAM (2.26) consists, in the general case, of the intrinsic and extrinsic contributions [28–32]. Indeed, a shift of the reference point $\mathbf{R}\to\mathbf{R}-\mathbf{R}_0$ produces change in the OAM: $\mathbf{J}_O \to \mathbf{J}_O - \mathbf{R}_0\times\mathbf{P}$. The natural reference point associated with the field itself is the field centroid (the energy-weighted mean position also known as 'centre of gravity', 'centre of energy', or 'moment of energy'):

$$\mathbf{R}_\text{c} = \frac{1}{W}\int \mathbf{R} w\, d^2\mathbf{r}\,. \tag{2.27}$$

Therefore, the intrinsic (i.e., origin-independent) and extrinsic parts of the OAM can be separated as [23,28,29,32]

$$\mathbf{J}_O^\text{int} = \mathbf{J}_O - \mathbf{R}_\text{c}\times\mathbf{P}\,,\quad \mathbf{J}_O^\text{ext} = \mathbf{R}_\text{c}\times\mathbf{P}\,. \tag{2.28}$$

Thus, the extrinsic OAM of the field is related to the evolution of the whole beam as a classical point particle with coordinate $\mathbf{R}_\text{c}$ and momentum $\mathbf{P}$. In other words, the extrinsic OAM is associated with the geometrical-optics trajectory of the beam centroid, while the intrinsic OAM describes energy flows taken with respect to the centre of the field (e.g., vortices).

The spin and orbital parts, (2.25) and (2.26), of the AM of the field are distinctly associated with the corresponding parts of the energy flow (the momentum density): SFD (2.8), (2.20) and OFD (2.9), (2.21). At the same time, the extrinsic part of the OAM can be associated with the total momentum of the field (2.23).

This division classifies the energy flows with respect to the 'physical nature' of different degrees of freedom: in a similar manner, for a moving atom we distinguish the 'extrinsic' motion of the atom 'as a whole' and the 'intrinsic' motion of the atomic electrons with their spin and orbital degrees of freedom. For further references, it would be suitable to call this classification

'structural', just in contrast to separation of contributions of the orthogonally polarized components (see section 3 below). Interactions and mutual transformations of the three different forms of the AM, as well as of corresponding energy flow contributions, constitute a very interesting and rapidly developing branch of optics [24,33–70]; some examples will be touched upon in subsequent sections.

## 3. Energy flows in paraxial beams

### *3.1. Basic properties*

Paraxial light beams represent the most important (both for fundamental theory and practical application) configurations of optical fields. In the paraxial approximation of Maxwell equations [21], the beam field can be expressed as a superposition of orthogonally polarized components characterized by the slowly varying complex amplitudes $u_\sigma(\mathbf{r}, z)$ which obey the equation

$$i\frac{\partial u_\sigma}{\partial z} = -\frac{1}{2k}\nabla_\perp^2 u_\sigma . \qquad (3.1)$$

Here $\nabla_\perp = \mathbf{e}_x\left(\partial/\partial x\right) + \mathbf{e}_y\left(\partial/\partial y\right)$ is the transverse gradient, whereas $\sigma = \pm 1$ for the basis of circular polarizations or $\sigma = x, y$ for the basis of linear polarization, which is equally admissible in the paraxial approximation. The complex amplitudes $u(\mathbf{r}, z)$ can be represented via real amplitude $A(\mathbf{r}, z)$ and phase $\varphi(\mathbf{r}, z)$:

$$u_\sigma = A_\sigma \exp\left(i\varphi_\sigma\right). \qquad (3.2)$$

The vector complex amplitude of the field is given by

$$\mathbf{u} = \mathbf{e}_x u_x + \mathbf{e}_y u_y = \mathbf{e}_+ u_+ + \mathbf{e}_- u_- , \qquad (3.3)$$

where

$$\mathbf{e}_+ = \frac{1}{\sqrt{2}}\left(\mathbf{e}_x + i\mathbf{e}_y\right), \quad \mathbf{e}_- = \frac{1}{\sqrt{2}}\left(\mathbf{e}_x - i\mathbf{e}_y\right) \qquad (3.4)$$

appear as a paraxial version of the helicity basis (2.15), (2.16), which does not depend on $\mathbf{k}$ ($\mathbf{e}_\mathbf{k} \simeq \mathbf{e}_z , \mathbf{e}_0 = -\mathbf{e}_x$). In terms of the complex amplitude $\mathbf{u}$, the paraxial electric and magnetic field strengths read [22]

$$\mathbf{E} = \mathbf{E}_\perp + \mathbf{e}_z E_z = \left[\mathbf{u} + \frac{i}{k}\mathbf{e}_z\left(\nabla_\perp \cdot \mathbf{u}\right)\right]e^{ikz}, \qquad (3.5)$$

$$\mathbf{H} = \mathbf{H}_\perp + \mathbf{e}_z H_z = \left[\left(\mathbf{e}_z \times \mathbf{u}\right) + \frac{i}{k}\mathbf{e}_z\left(\nabla_\perp \cdot \left(\mathbf{e}_z \times \mathbf{u}\right)\right)\right]e^{ikz}. \qquad (3.6)$$

The main (first) terms of (3.5) and (3.6) describe the transverse field components, whereas the longitudinal components (the second terms) are of the relative order $\gamma = \left(kb\right)^{-1}$ in magnitude, with $b$ being the characteristic transverse scale of the distribution $\mathbf{u}(\mathbf{r}, z)$. The quantity $\gamma$ is the small parameter of the paraxial approximation.

In the first-order approximation in $\gamma$, the longitudinal field does not affect the energy density (2.2) which takes the form

$$w = g\left(\mathbf{u}^* \cdot \mathbf{u}\right) = g\sum_\sigma A_\sigma^2 \equiv \sum_\sigma w_\sigma . \qquad (3.7)$$

The paraxial version of 'spin-orbit' decomposition (2.7) – (2.9) of the energy flow reads

$$\mathbf{p}_S = -\frac{i}{2\omega}g\left(\nabla_\perp \times \left[\mathbf{u}^* \times \mathbf{u}\right]\right), \quad \mathbf{p}_O = \frac{g}{c}\left[\mathbf{e}_z\left(\mathbf{u}^* \cdot \mathbf{u}\right) + \frac{1}{k}\mathrm{Im}\left(\mathbf{u}^* \cdot \left(\nabla_\perp\right)\mathbf{u}\right)\right]. \qquad (3.8)$$

While the spin flow represented by the SFD distribution $\mathbf{p}_S$ is purely transverse, the OFD $\mathbf{p}_O$ has the longitudinal component

$$p_{Oz} = p_z = \frac{w}{c}. \tag{3.9}$$

Equation (3.9) reproduces the energy-momentum relation typical for a plane wave or for a photon [14]. The intensity of the beam (e.g., in Watts per unit area) is just the longitudinal energy flow density component that due to (2.5) and (3.9) is given by

$$I = c^2 p_z = cw \tag{3.10}$$

– in paraxial case the beam intensity profile $I(\mathbf{r})$ can be equally characterized by functions $w(\mathbf{r})$ or $p_z(\mathbf{r})$. In practice, different detectors measure either $w$ or the energy flow within a limited solid angle (approximation of $p_z$) [71,72]; in paraxial conditions, due to relation (3.10), they are equivalent.

The longitudinal flow (3.9) makes no contribution to the beam AM with respect to the propagation axis $z$. The second term in the second equation (3.8) describes the transverse part of the OFD, which, using (3.2) and (3.3), can be written as

$$\mathbf{p}_{O\perp} = \frac{g}{\omega}\sum_\sigma A_\sigma^2 \nabla_\perp \varphi_\sigma = \frac{1}{\omega}\sum_\sigma w_\sigma \nabla_\perp \varphi_\sigma . \tag{3.11}$$

Linear densities of the energy and momentum per unit $z$-length are given by

$$W = g\int \left(\mathbf{u}^* \cdot \mathbf{u}\right) d^2\mathbf{r}, \quad P_z = \frac{W}{c}, \tag{3.12}$$

$$\mathbf{P}_\perp = \frac{g}{\omega}\operatorname{Im}\int \left(\mathbf{u}^* \cdot (\nabla_\perp)\mathbf{u}\right) d^2\mathbf{r} = \frac{g}{\omega}\sum_\sigma \int A_\sigma^2 \nabla_\perp \varphi_\sigma d^2\mathbf{r} . \tag{3.13}$$

Note that the spin flow makes no contribution to the transverse momentum (3.13). Akin to (3.10), the experimentally-measured power of the beam (e.g., in Watts) is determined by the total longitudinal momentum, that is

$$\Phi = c^2 P_z = cW . \tag{3.14}$$

The SAM and OAM with respect to the $z$-axis per unit $z$-length of the beam read

$$\mathbf{J}_S = \int \mathbf{r} \times \mathbf{p}_S \, d^2\mathbf{r} = \mathbf{e}_z J_S, \quad \mathbf{J}_O = \int \mathbf{r} \times \mathbf{p}_O \, d^2\mathbf{r} = \mathbf{e}_z J_O . \tag{3.15}$$

Calculations with equations (3.8) and (3.11) result in

$$J_S = \frac{i}{2\omega} g \int \mathbf{r} \cdot \nabla \left(\mathbf{u}^* \times \mathbf{u}\right)_z d^2\mathbf{r} = -\frac{i}{\omega} g \int \left(\mathbf{u}^* \times \mathbf{u}\right)_z d^2\mathbf{r}, \tag{3.16}$$

$$J_O = \frac{g}{\omega}\sum_\sigma \int A_\sigma^2 \left(\mathbf{r} \times \nabla \varphi_\sigma\right)_z d^2\mathbf{r} = \frac{g}{\omega}\sum_\sigma \int A_\sigma^2(\mathbf{r}) \frac{\partial \varphi_\sigma(\mathbf{r})}{\partial \phi} d^2\mathbf{r}, \tag{3.17}$$

where $\phi = \arctan(y/x)$ is the azimuthal angle in the beam cross section.

Equations (3.12) – (3.17) are paraxial counterparts of relations (2.22) – (2.26), which are written in the coordinate representation. In the plane-wave (angular spectrum) representation, the paraxial regime is characterized by almost longitudinal wave vectors: $\mathbf{k} = k\mathbf{e}_z + \mathbf{k}_\perp$, $|\mathbf{k}_\perp| \ll k$, and nearly transverse fields $\mathbf{E}$ and $\mathbf{H}$. Any plane-wave component propagates in direction which makes with the $z$ axis small angles $q_x$ and $q_y$ that form a dimensionless vector $\mathbf{q} = \mathbf{k}_\perp / k \sim \gamma$. In the helicity basis (3.4)

$$u_\sigma(\mathbf{R}) \simeq \frac{1}{2\pi}\int C_\sigma(\mathbf{k}) e^{i\mathbf{k}\cdot\mathbf{R} - ikz} d^2\mathbf{k} . \tag{3.18}$$

where $d^2\mathbf{k} \simeq dk_x dk_y = k^2 dq_x dq_y$, integration is taken over the whole $(k_x, k_y)$-plane, $\sigma = \pm 1$ and $k_z = \sqrt{k^2 - k_\perp^2} \simeq k - k_\perp^2/2k$. This results in

$$W = g\sum_\sigma \int_{|\mathbf{k}|=k} \left|C_\sigma(\mathbf{k})\right|^2 d^2\mathbf{k} \equiv \sum_\sigma W_\sigma, \quad P_z = \frac{W}{c}, \tag{3.19}$$

$$\mathbf{P}_\perp = \frac{g}{\omega}\sum_\sigma \int_{|\mathbf{k}|=k} \mathbf{k}_\perp \left|C_\sigma(\mathbf{k})\right|^2 d^2\mathbf{k} \tag{3.20}$$

$$J_S = \frac{g}{\omega}\sum_\sigma \int_{|\mathbf{k}|=k} \sigma\left|C_\sigma(\mathbf{k})\right|^2 d^2\mathbf{k}, \tag{3.21}$$

$$J_O = \frac{g}{\omega}\sum_\sigma \int_{|\mathbf{k}|=k} C_\sigma^*(\mathbf{k})\cdot\left(-i\frac{\partial}{\partial\tilde{\phi}}\right)C_\sigma(\mathbf{k})d^2\mathbf{k}, \tag{3.22}$$

where $\tilde{\phi} = \arctan(k_y/k_x)$ is the azimuthal angle in the $\mathbf{k}$-space. Equations (3.19) – (3.22) are equivalent to corresponding equations (3.12), (3.13), (3.16) and (3.17) owing to (3.2) – (3.4) and (3.18). From equations (3.19) and (3.21) it follows that the 'normalized' SAM (i.e., the 'SAM per photon') [4] is $J_S\omega/W = \bar{\sigma}$, where $\bar{\sigma} = (W_+ - W_-)/(W_+ + W_-)$ is the average helicity. At the same time, for circularly symmetric optical-vortex fields

$$A_\sigma(\mathbf{r}, z) = A_\sigma(r, z), \quad \varphi_\sigma(\mathbf{r}, z) = l\phi + f(r, z), \tag{3.23}$$

$C_\sigma \propto \exp(il\tilde{\phi})$, and the OAM per photon is equal to $J_O\omega/W = l$ [1–4]. Here $l$ is an integer (for $l \neq 0$, conditions (3.23) specify standard models of the optical-vortex beam [1–4,73], see equations (4.2) and (4.6) below).

According to (2.28), the transverse momentum $\mathbf{P}_\perp$ (3.13), (3.20) together with the transverse displacement of the beam centroid (2.27),

$$\mathbf{r}_c = \frac{1}{W}\int \mathbf{r} w d^2\mathbf{r}, \tag{3.24}$$

give the extrinsic contribution to the longitudinal OAM

$$J_O^{\text{ext}} = (\mathbf{r}_c \times \mathbf{P}_\perp)_z, \quad J_O^{\text{int}} = J_O - (\mathbf{r}_c \times \mathbf{P}_\perp)_z. \tag{3.25}$$

These equations show that the extrinsic OAM is produced by the tilt of the beam propagation direction, $\mathbf{P}_\perp \neq 0$, and simultaneous orthogonal displacement of the beam centre, $\mathbf{r}_c \neq 0$ [28–32]. Such interrelated tilts and displacements appear in some beam transformations resulting in conversions between intrinsic SAM or OAM and extrinsic OAM of the beam [24,42,46,47]. For instance, similar transformations can be caused just by a rotation of the coordinate frame which also generates non-zero transverse extrinsic OAM [46] (see section 6.2 below).

### *3.2. Internal flows and the irradiance moments of the beam*

The AM of the beam can be considered as an integral characteristic of the transverse energy flow, especially, of its circulation components. For a scalar paraxial field with complex amplitude $u(\mathbf{r})$, similar characteristics can be introduced that are associated with the beam Wigner function defined by equation [74,75]

$$G(\mathbf{r}, \mathbf{q}) = \frac{k^2}{4\pi^2}\int u\left(\mathbf{r} + \frac{\mathbf{r}'}{2}\right)u^*\left(\mathbf{r} - \frac{\mathbf{r}'}{2}\right)\exp\left[-ik(\mathbf{q}\cdot\mathbf{r}')\right]d^2\mathbf{r}', \tag{3.26}$$

where vector $\mathbf{q} = (q_x, q_y)$ specifies the direction of a plane-wave component (see the note before equation (3.18)). The Wigner function (3.26) is always real, and the energy density in the beam (3.7) can be found as

$$w(\mathbf{r}) = g\int G(\mathbf{r},\mathbf{q})\,d^2\mathbf{q}\,. \tag{3.27}$$

Accordingly, the linear energy density (3.12) is

$$W = g\int G(\mathbf{r},\mathbf{q})\,d^2\mathbf{r}\,d^2\mathbf{q}\,. \tag{3.28}$$

The most useful applications of the Wigner function are associated with its moments – 'irradiance moments' of the beam. They form the basis of fruitful tools for parametric characterization of the laser beams which are adopted as ISO standards [76]. The most important are the first and second moments. The first moments form the 4-vector

$$\mathbf{r}_c = \begin{pmatrix} x_c \\ y_c \end{pmatrix} = \frac{g}{W}\int \mathbf{r}G(\mathbf{r},\mathbf{q},z)\,d^2\mathbf{r}\,d^2\mathbf{q}\,, \tag{3.29}$$

$$\mathbf{q}_c = \begin{pmatrix} q_{xc} \\ q_{yc} \end{pmatrix} = \frac{g}{W}\int \mathbf{q}G(\mathbf{r},\mathbf{q},z)\,d^2\mathbf{r}\,d^2\mathbf{q} = \frac{\mathbf{P}_\perp}{P_z}\,, \tag{3.30}$$

which unites the spatial and angular coordinates of the centroid trajectory. It can be easily seen that, for scalar beams, equations (3.29) and (3.30) yield the same results as (3.24) and the combination of (3.13) and the second equation (3.12), so that $\mathbf{r}_c$ (3.29) and $\mathbf{q}_c$ (3.30) characterize the mean transverse position and the mean transverse momentum of the beam [74,75].

In case of a vector paraxial beam, one can introduce these quantities for a single scalar $\sigma$-component, and relations (3.29) and (3.30) reduce to

$$\mathbf{r}_{c\sigma} = \frac{\int \mathbf{r}\left|u_\sigma\right|^2 d^2\mathbf{r}}{\int \left|u_\sigma\right|^2 d^2\mathbf{r}}\,, \quad \mathbf{q}_{c\sigma} = -\frac{i}{k}\frac{\int u_\sigma^* \nabla_\perp u_\sigma d^2\mathbf{r}}{\int \left|u_\sigma\right|^2 d^2\mathbf{r}} = \frac{\mathbf{P}_{\sigma\perp}}{P_{\sigma z}}\,.$$

The corresponding centroid characteristics of the whole vector beam can be found as weighted averages of the two polarization components:

$$\mathbf{r}_c = \frac{1}{W}\sum_\sigma \mathbf{r}_{c\sigma}W_\sigma\,, \quad \mathbf{q}_c = \frac{1}{W}\sum_\sigma \mathbf{q}_{c\sigma}W_\sigma\,. \tag{3.31}$$

The second irradiance moments form the symmetric positive-definite 4×4 'irradiance moment matrix' [74,75] which can be represented via 2×2 blocks:

$$\mathsf{M} = \begin{pmatrix} \mathsf{M}_{11} & \mathsf{M}_{12} \\ \tilde{\mathsf{M}}_{12} & \mathsf{M}_{22} \end{pmatrix},$$

where the tilde stands for the matrix transposition. The transverse flow characteristics are 'enclosed' in the off-diagonal block

$$\mathsf{M}_{12} \equiv \begin{pmatrix} m_{xx} & m_{xy} \\ m_{yx} & m_{yy} \end{pmatrix} = \frac{g}{W}\int \begin{pmatrix} xq_x & xq_y \\ yq_x & yq_y \end{pmatrix} G(\mathbf{r},\mathbf{q})\,d^2\mathbf{r}\,d^2\mathbf{q} = \frac{c}{W}\int \begin{pmatrix} xp_x & xp_y \\ yp_x & yp_y \end{pmatrix} d^2\mathbf{r}\,. \tag{3.32}$$

Due to the last equation (3.32), elements of matrix $\mathsf{M}_{12}$ are related to the transverse components of the beam momentum density. As immediately measurable parameters, in some cases they give access to experimental investigation of the transverse energy flows (see also section 6.2 below). In particular, the beam OAM can be determined as [77–79]

$$J_O = \frac{W}{c}\left(m_{xy} - m_{yx}\right). \tag{3.33}$$

Relations (3.32), (3.33) enable one to operate with the average characteristics of the beam flows by means of the well elaborated general scheme of the beam characterization [75,76]. Some important characteristics of the internal energy flows in paraxial beams can be suitably described with the help of the irradiance moments and related quantities. In particular, representation (3.33) makes it possible to consider the energy flows and the OAM of partially coherent beams [80], to introduce conceptions of "vortex" and "asymmetry" OAM [77,79] and to describe the AM transformations in the first-order optical systems [75,77,80–82].

*3.3. Spin flow*

The SFD (3.8) can be expressed in the form

$$\mathbf{p}_S = \frac{1}{2\omega c}\left(\mathbf{e}_x \frac{\partial s_3}{\partial y} - \mathbf{e}_y \frac{\partial s_3}{\partial x}\right) = -\frac{1}{2\omega c}\left[\mathbf{e}_z \times \nabla_\perp s_3\right] \qquad (3.34)$$

where

$$s_3 \equiv s_3(\mathbf{r}, z) = -icg\mathbf{e}_z \cdot \left(\mathbf{u}^* \times \mathbf{u}\right) = icg\left(u_X u_Y^* - u_X^* u_Y\right) = I_+ - I_- \qquad (3.35)$$

is the spatial distribution of the 'third' Stokes parameter characterizing the degree of circular polarization [14], $I_\sigma$ is the partial intensity of a single polarization component defined like (3.10). For beams with homogeneous polarization whose complex amplitude vector (3.3) can be represented as $\left(\alpha\mathbf{e}_x + \beta\mathbf{e}_y\right)u$ with scalar $u \equiv u(\mathbf{r}, z)$ [1], ratio $s_3/I$ is equivalent to the frequently used ellipticity parameter $i\left(\alpha\beta^* - \alpha^*\beta\right)$ (see, e.g., [1,23,25,46]). Due to (3.35), each circularly polarized component possesses its own partial spin flow:

$$\mathbf{p}_S = \mathbf{p}_{+S} + \mathbf{p}_{-S}, \quad \mathbf{p}_{\sigma S} = -\frac{\sigma}{2\omega c}\left[\mathbf{e}_z \times \nabla_\perp I_\sigma\right] = \frac{\sigma}{2\omega c}\nabla_\perp \times \left(\mathbf{e}_z I_\sigma\right) \quad (\sigma = \pm 1). \qquad (3.36)$$

In particular, in polar coordinates

$$\mathbf{p}_{\sigma S} = -\frac{\sigma}{2\omega c}\left(-\mathbf{e}_r \frac{1}{r}\frac{\partial}{\partial\phi} + \mathbf{e}_\phi \frac{\partial}{\partial r}\right) I_\sigma(r, \phi) \qquad (3.37)$$

where $r = |\mathbf{r}|$, and unit vectors of the polar coordinates

$$\mathbf{e}_r = \frac{\mathbf{e}_x x + \mathbf{e}_y y}{r}, \quad \mathbf{e}_\phi = \frac{-\mathbf{e}_x y + \mathbf{e}_y x}{r} \qquad (3.38)$$

are introduced. In paraxial beams the SFD is a transverse 2D vector field (no longitudinal component). As curls of certain vector fields the SFD itself and its partial contributions preserve the 2D version of the solenoidal property (2.11) so the spin flow lines are everywhere continuous.

Due to (3.34) – (3.36), the SFD is closely associated with the $s_3(\mathbf{r}, z)$ distribution. The spin flow lines coincide with the constant-level lines of $s_3(\mathbf{r})$; when moving along such a line following the flow direction, the area of high $s_3$ always remains to the left (figure 1). In particular, near extrema of $s_3(\mathbf{r})$ the SFD possesses a circulatory character (see figure 1b) with circulation determined by the paraxial version of vorticity (2.12)

$$\mathbf{\Omega}_S = \nabla_\perp \times \mathbf{p}_S = -\frac{g}{2\omega}\mathrm{Im}\left[\nabla_\perp^2\left(\mathbf{u}^* \times \mathbf{u}\right)\right] = -\mathbf{e}_z \frac{1}{2k}\nabla_\perp^2 s_3 = \mathbf{e}_z \frac{1}{2k}\nabla_\perp^2\left(I_- - I_+\right). \qquad (3.39)$$

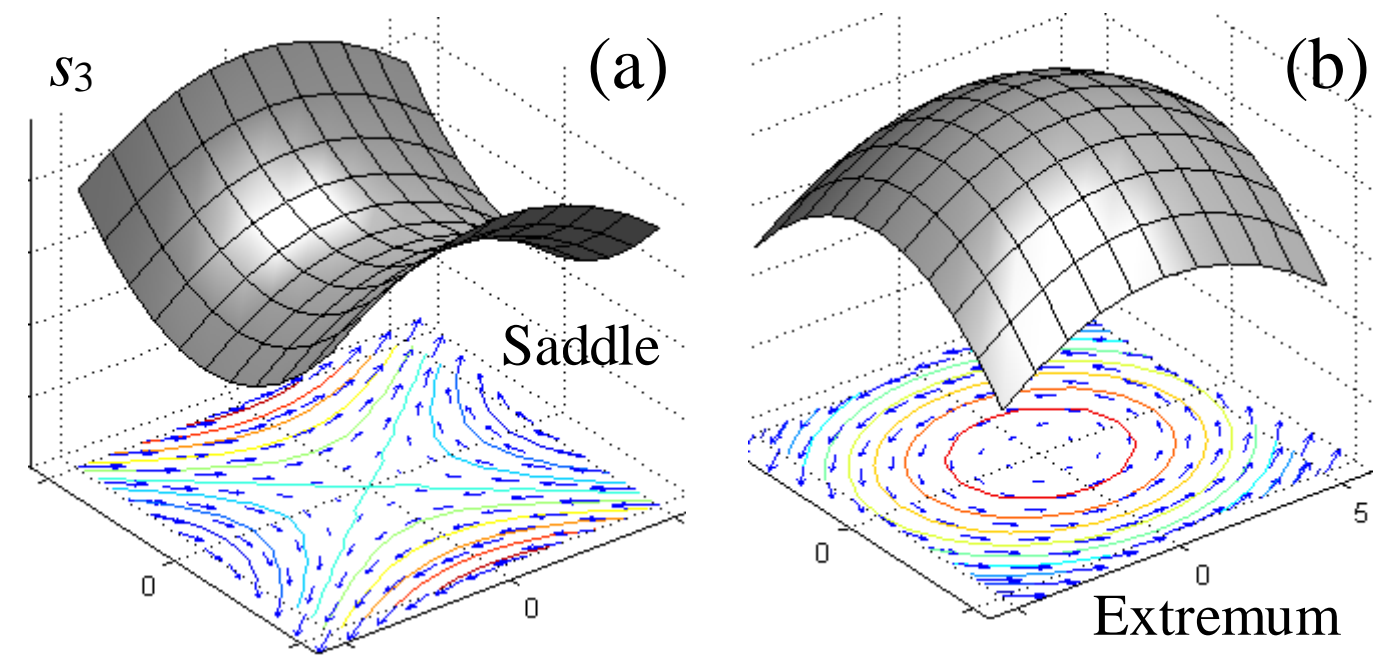

**Figure 1**. The SFD vector lines near the (a) saddle and (b) maximum of the $s_3(\mathbf{r})$ distribution.

This behaviour can be explained by the simple phenomenological model of the spin flow in paraxial beams [4,83] schematically illustrated in figure 2a. The physical ground for the energy circulation in the circularly polarized beams is the rotation of the field vectors that takes place ‘in every point’ of the field. One can imagine that situation looks as if the energy circulates within microscopic “cells”; if the cells are identical, contributions of the adjacent cells compensate each other and the macroscopic energy flow is absent. The compensation is not complete if the adjacent cells differ (the beam is transversely inhomogeneous), and this explains why the SFD is orthogonal to the inhomogeneity gradient (figure 2b). The compensation completely disappears if the cell series breaks, i.e. at the beam boundary. This must not obligatory be a real physical boundary; no matter how a certain part of the beam cross section is isolated, its near-boundary cells will be ‘uncompensated’ and the resulting energy circulation will appear along this boundary (see figure 2a). Therefore, the SAM of any fragment $A$ of the beam cross section contains not only the ‘bulk’ contribution given by (3.16) but also the contribution of the boundary $F_A$ [83]:

$$J_S(A) = \frac{1}{2\omega c} \oint_{F_A} s_3(\mathbf{r}) |\mathbf{r} \times d\mathbf{r}| . \qquad (3.40)$$

Adding this to (3.16) and allowing for (3.34) and (3.35), we arrive at the universal SAM expression correct for the whole beam as well as for its arbitrary transverse fragment

$$J_S(Q) = \int |\mathbf{r} \times \mathbf{p}_S| d^2\mathbf{r} + \frac{1}{2\omega c} \oint_A s_3(\mathbf{r}) |\mathbf{r} \times d\mathbf{r}| = \frac{1}{2\omega c} \left[ -\int_Q \mathbf{r} \cdot \nabla_\perp s_3(\mathbf{r}) d^2\mathbf{r} + \oint_A s_3(\mathbf{r}) |\mathbf{r} \times d\mathbf{r}| \right]. \qquad (3.41)$$

After integration by parts it can be represented in the usual form

$$J_S(Q) = \int_Q J'_S(\mathbf{r}) d^2\mathbf{r} \qquad (3.42)$$

where

$$J'_S(\mathbf{r}) = -\frac{i}{\omega} g |\mathbf{u}^* \times \mathbf{u}| = \frac{1}{\omega c} s_3(\mathbf{r}) \qquad (3.43)$$

is the SAM volume density (cf. second equation (3.16)). For usual transversely limited beams with smooth intensity fall-off, this conclusion fully coincides with the known result [73]. Remarkably, due to explicit introduction of the boundary contribution (3.40) the known paradoxes associated with the SAM of transversely limited light beams [4,84–86] appear to be eliminated. An analogous but mathematically more exquisite approach was employed by A.M. Stewart [87].

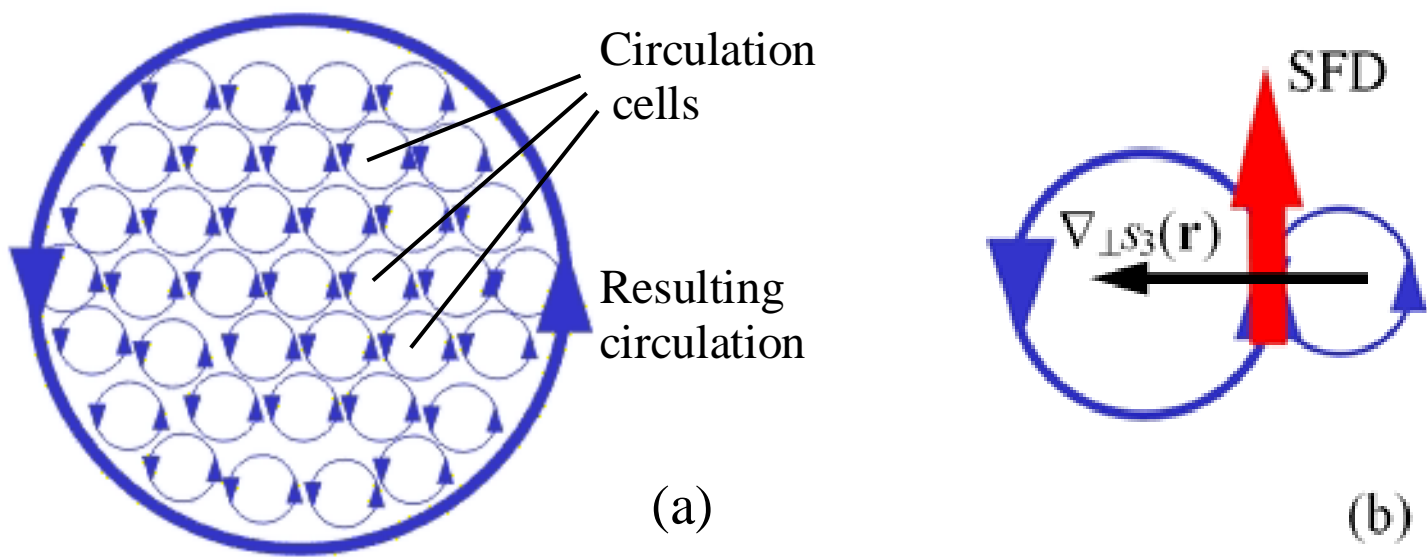

**Figure 2**. Model pattern of the spin flow within the cross section of a beam with circular polarization: (a) homogeneous beam with abrupt boundary, (b) emergence of the macroscopic spin flow (red arrow) when the circulation cells do not compensate (inhomogeneous beam: cell sizes conventionally reflect the “strengths” of adjacent circulations). Black arrow shows the inhomogeneity gradient.

## *3.4. Orbital flow*

In accordance with (3.3) and (3.17), the OFD of a paraxial beam equals to the sum of analytically identical contributions of the two orthogonal polarizations, no matter linear or circular,

$$\mathbf{p}_{O\perp} = \sum_\sigma \mathbf{p}_{\sigma O}\,, \quad \mathbf{p}_{\sigma O} = \frac{g}{\omega}\operatorname{Im}\left(u_\sigma^* \nabla_\perp u_\sigma\right) = \frac{1}{\omega c} I_\sigma \nabla_\perp \varphi_\sigma \quad (\sigma = X,\, Y, \text{ or } +1,\, -1). \tag{3.44}$$

A more explicit expression in the polar coordinates will be useful for further references:

$$\mathbf{p}_{\sigma O} = \frac{1}{\omega c} I_\sigma \nabla \varphi_\sigma = \frac{1}{\omega c} I_\sigma \left( \mathbf{e}_\phi \frac{1}{r}\frac{\partial}{\partial \phi} + \mathbf{e}_r \frac{\partial}{\partial r} \right) \varphi_\sigma\,. \tag{3.45}$$

Partial OFD fields $\mathbf{p}_{\sigma O}$ are determined by the corresponding phase distributions $\varphi_\sigma(\mathbf{r})$. In fact, the transverse orbital flow is formed by transverse projections of rays that are orthogonal to the partial wavefronts. The orbital flow pattern is associated with the relief of function $\varphi_j(\mathbf{r})$ similarly to how the spin flow is associated with the relief of $s_3(\mathbf{r})$ (figure 1) but the OFD lines are orthogonal to the contours of constant phase (figure 3). In contrast to the total orbital flow (2.9), its transverse part (3.44) does not obey the solenoidal field condition (2.11); the vector lines are not continuous in figures 3b, 3c. Moreover, the transverse flow divergence plays important role in the beam transformation during its propagation. This follows from the 'continuity equation'

$$\frac{\partial w}{\partial z} = -c\left(\nabla_\perp \mathbf{p}_\perp\right) = -c\left(\nabla_\perp \mathbf{p}_{O\perp}\right) \tag{3.46}$$

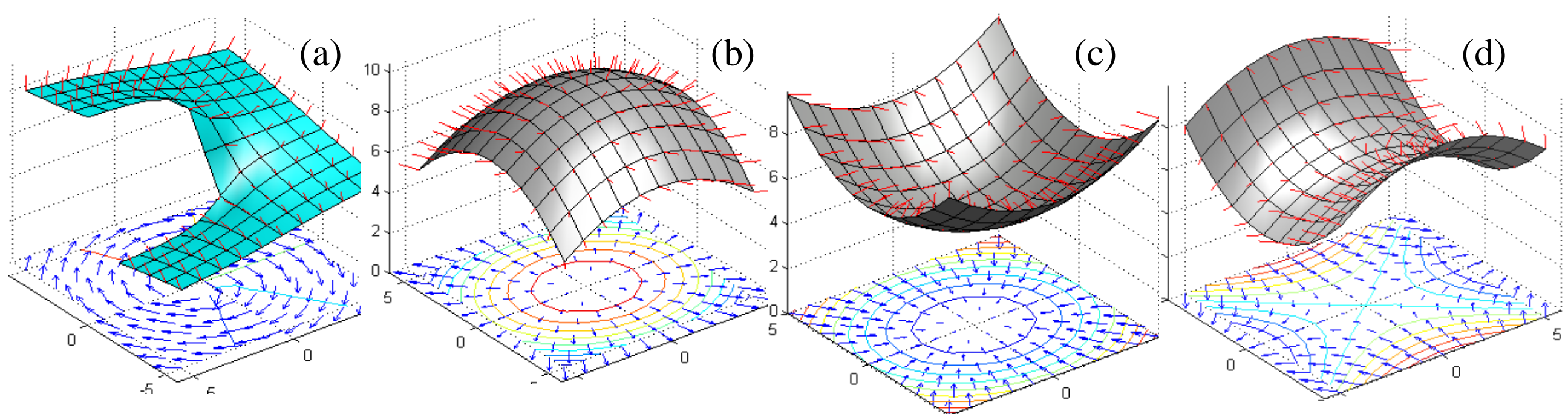

**Figure 3**. Patterns of the partial OFDs with corresponding wavefronts: (a) phase vortex, OFD is circulatory; (b) phase minimum, wavefront is convex, OFD is divergent; (c) phase maximum, wavefront is concave, OFD is convergent; (d) phase and wavefront saddle.

that can be easily derived from (3.1), (3.9) and (3.10) – (3.12). In accordance with (3.46), it is the OFD that is responsible for the well known phenomena of the beam divergence, self-diffraction and transverse energy circulation. The latter is associated with the azimuthal OFD component that, due to (3.17), leads to the orbital AM of the beam. The general measure of the OFD circulation is provided by the corresponding 'orbital' vorticity

$$\mathbf{\Omega}_O = \nabla_\perp \times \mathbf{p}_O = \frac{g}{2\omega}\sum_\sigma \operatorname{Im}\left(\nabla_\perp u_\sigma^* \times \nabla_\perp u_\sigma\right). \tag{3.47}$$

Due to its solenoidal character, the SFD drops out of equation (3.46): it is important that the spin flow does not affect the variations of intensity of a freely propagating paraxial beam.

Confrontation of equations (3.34), (3.36) and (3.39) on the one hand, and (3.44), (3.47) on the other, demonstrates main formal differences between the spin and orbital flow descriptions. For the spin flow characteristics, the partial contributions of different polarizations can be separated only in the helicity basis (3.4) while the OFD characteristics are separable in any orthogonal polarization basis. This is connected with another important difference: partial contributions to the OFD add together while the partial SFD contributions are combined with opposite signs. The fact that a linear polarization has 'no sign' agrees with that a linearly polarized beam has no SFD. A linearly polarized beam can be represented as a sum of the left- and right-polarized beams which are identical in all other respects, so their SFDs completely cancel each other. The OFD expressions (3.44) and (3.47) are invariant in respect of unitary transformations of the transverse coordinates; all partial contributions possess identical analytical structures following from (3.44) and (3.47), and we can study one of them thus reducing the problem to a scalar case. That is why scalar field models are so popular in connection to the OAM and related issues [1–4], and in subsequent analysis of the OFD we will consider the partial contributions separately in all cases where this is possible, omitting the polarization index $\sigma$.

*3.5. Internal flows and the instant pattern rotation*

It is commonly recognized that the spin flow originates from the optical-frequency rotation of the field vectors: despite that this rotation has no mechanical meaning, it is a source of real energy circulation and real AM of light. Likewise, the circulatory behaviour of the OFD is associated with the rotational component in the pattern of the instant field oscillations within the beam cross section [88]. For the scalar model, the instant electric field in a given cross section ($z$ = const) follows from (3.2), (3.5) and (2.1) in the form

$$E(\mathbf{r},t) = A(\mathbf{r})\cos\left[\varphi(\mathbf{r}) - \omega t\right]. \tag{3.48}$$

For example, in the important case of a circularly symmetric optical-vortex beam (see (3.23)), the instant field rotates with frequency $\omega/l$ [2]:

$$E(\mathbf{r},t) = A(r)\cos(l\phi - \omega t). \tag{3.49}$$

This rotation is a sort of ‘sunlight spot’ motion and has no direct mechanical meaning – any point of distribution (3.49) lying at a distance > $l\lambda/2\pi$ from the axis moves with velocity > $c$. However, it is directly linked to the real OAM of the beam.

Note that in the case of ‘pure’ rotation (3.49), the time and azimuth derivatives of the instant field distribution are proportional,

$$l\frac{\partial E(\mathbf{r},t)}{\partial \phi} = -\omega\frac{\partial E(\mathbf{r},t)}{\partial t}.$$

It can be expected that in more complicated situations of arbitrary field (3.48), the presence of rotational component in the whole pattern of the instant field oscillation can be manifested in certain correlations between these quantities. The usual measure of such correlations is the correlation coefficient

$$\left\langle \frac{\partial E(\mathbf{r},t)}{\partial \phi}\frac{\partial E(\mathbf{r},t)}{\partial t}\right\rangle \propto K = \int \frac{\partial E(\mathbf{r},t)}{\partial \phi}\frac{\partial E(\mathbf{r},t)}{\partial t} d^2\mathbf{r}\, dt \tag{3.50}$$

where the integration performs averaging over the beam cross section and the oscillation period. After substitution of (3.48) and comparing the result with a summand of (3.17) one readily obtains

$$K = -\pi\frac{\omega}{g}J_O. \tag{3.51}$$

The correlation integral (3.50) that testifies for the rotational component in the instantaneous oscillatory pattern (3.48) is proportional to the usual measure of the OAM in a scalar beam. Therefore, in any beam where the transverse OFD has a circulatory character, the instantaneous field distribution shows a sort of rotation within the beam cross section. The natural characteristic of this rotation is the beam OAM which, in case of a scalar optical field, is not only a witness but also a direct kinematic measure of the presence of the rotational component in the whole pattern of the instant oscillations [88]. Just like the spin flow owes to the rotation of the instant field vectors that takes place in every point of the beam cross section, the circulatory orbital flow is associated with the rotational behaviour of the instant field distribution around the beam axis. This difference is inherent in all manifestations of the spin and orbital flows and can serve to physically distinguish the two forms of the light ‘rotation’.

In a similar way, the radial OFD component (second term in parentheses of (3.45)) can be related to the radial (centrifugal or centripetal) ‘motion’ of the instantaneous field pattern. This provides sound arguments that any transverse energy flow in a light beam is associated with the corresponding kinematic behaviour of the instantaneous field in the beam cross section.

## 4. Model patterns of the energy flow

The energy flow lines in an arbitrary electromagnetic field are determined by the differential equation [17,89]

$$\frac{d\mathbf{R}}{d\tau} = \mathbf{p}(\mathbf{R}). \tag{4.1}$$

The sought lines $\mathbf{R} = \mathbf{R}(\tau)$ are parameterized by the scalar parameter $\tau$. In any point, the tangent to line defined via (4.1) is parallel to the local $\mathbf{p}(\mathbf{R})$. Patterns of the separate partial and structural energy flow components, discussed in the preceding sections, are determined in quite analogous way.

*4.1. Scalar beams with optical vortices*

In this section, we consider scalar beam models where the internal flow is of purely orbital nature. A simple and instructive example is provided by the circular Laguerre-Gaussian (LG) beams – the standard models of light fields with OAM [1–4]. Their properties, including the Poynting vector behaviour, subjected to study for a long time [73,90–92]. In this case, the complex amplitude distribution $u \equiv u_{ql}$ depends on two integer indices, arbitrary $l$ and positive $q$,

$$u_{ql} = \sqrt{\frac{8\Phi_{ql}}{c}} \sqrt{\frac{q!}{(q+|l|)!}} \frac{1}{b} \left(\frac{r}{b}\right)^{|l|} L_q^{|l|}\left(\frac{r^2}{b^2}\right) \exp\left(-\frac{r^2}{2b^2}\right) \exp\left[ik\frac{r^2}{2R} + il\phi - i\left(2q+|l|+1\right)\chi\right]. \tag{4.2}$$

$\Phi_{ql}$ is the power of the mode with indices $q$ and $l$, related to the beam energy per unit length $W_{ql}$ via (3.14), $L_q^{|l|}$ is the symbol of the Laguerre polynomial. Upon the beam propagation, its transverse radius $b$, radius of the wavefront curvature $R$ and the additional phase shift due to finite transverse beam size (Gouy phase [1–3]) $\chi$ change. If the beam waist is situated in the cross section $z = 0$, they obey the equations

$$R(z) = \frac{z_R^2 + z^2}{z}, \quad b^2(z) = \frac{z_R^2 + z^2}{kz_R}, \quad \chi(z) = \arctan\left(\frac{z}{z_R}\right), \tag{4.3}$$

where

$$z_R = kb_0^2 \tag{4.4}$$

is the confocal parameter (Rayleigh length) of the beam, $b_0$ being its radius in the waist cross section. With the help of (3.8), (3.9), (3.12) and (4.2), (4.3) one obtains the approximate explicit representation of the flow lines [90]

$$\mathbf{p}_O(\mathbf{r}, z) = \frac{w(\mathbf{r}, z)}{c}\left(\frac{zr}{z_R^2 + z^2}\mathbf{e}_r + \frac{l}{kr}\mathbf{e}_\phi + \mathbf{e}_z\right), \tag{4.5}$$

where

$$w(\mathbf{r}, z) = \frac{1}{c}\frac{\Phi_{ql}}{\pi b^2}\frac{q!}{(q+|l|)!}\left(\frac{r}{b}\right)^{2|l|}\left[L_q^{|l|}\left(\frac{r^2}{b^2}\right)\right]^2 \exp\left(-\frac{r^2}{b^2}\right).$$

Another example of the transverse energy circulation is associated with the Bessel beams [93,94]. In paraxial approximation they are described by the complex amplitude distribution

$$u = AJ_l\left(\frac{r}{b_0}\right)\exp\left(il\phi - i\frac{z}{2kb_0^2}\right), \tag{4.6}$$

where $A$ is the constant amplitude, $J_l$ is the Bessel function of the first kind, and $b_0^{-1} = k_\perp$ is the transverse radial wave-vector component which is constant for Bessel beams. Formally, the Bessel beam carries infinite power, since its envelope amplitude decays as $r^{-1/2}$ at large $r$. Nevertheless, the Bessel-beam solutions adequately describe nearly-diffractionless fields which can be generated in a finite region of space. The corresponding energy flow lines are given by the equation

$$\mathbf{p}_O(\mathbf{r}, z) = \frac{g}{c}|A|^2 J_l^2\left(\frac{r}{b_0}\right)\left(\frac{l}{kr}\mathbf{e}_\phi + \mathbf{e}_z\right), \tag{4.7}$$

which, in fact, is a simplified version of (4.5). A remarkable property of the Bessel beams is that they allow simple nonparaxial vector generalization. Spin and orbital energy flows in nonparaxial polarized beams essentially involve spin-orbit interaction and Berry-phase effects, which are examined in [24].

Flow lines calculated from relations (4.5) and (4.7) demonstrate the helical energy transport in freely propagating vortex beams, see figure 4. Due to the non-diffracting character, the transverse profile of the Bessel beam remains unchanged and the energy helixes lie on cylindrical surfaces (figure 4c,d). In contrast, the LG beams diverge and their energy flow streamlines have radial components (figure 4a,b). A comparative analysis of the energy flow streamlines and geometrical optics rays in the scalar Bessel and LG paraxial beams was carried out in [95]. In both cases the transverse velocity of the energy circulation is inversely proportional to the off-axial distance.

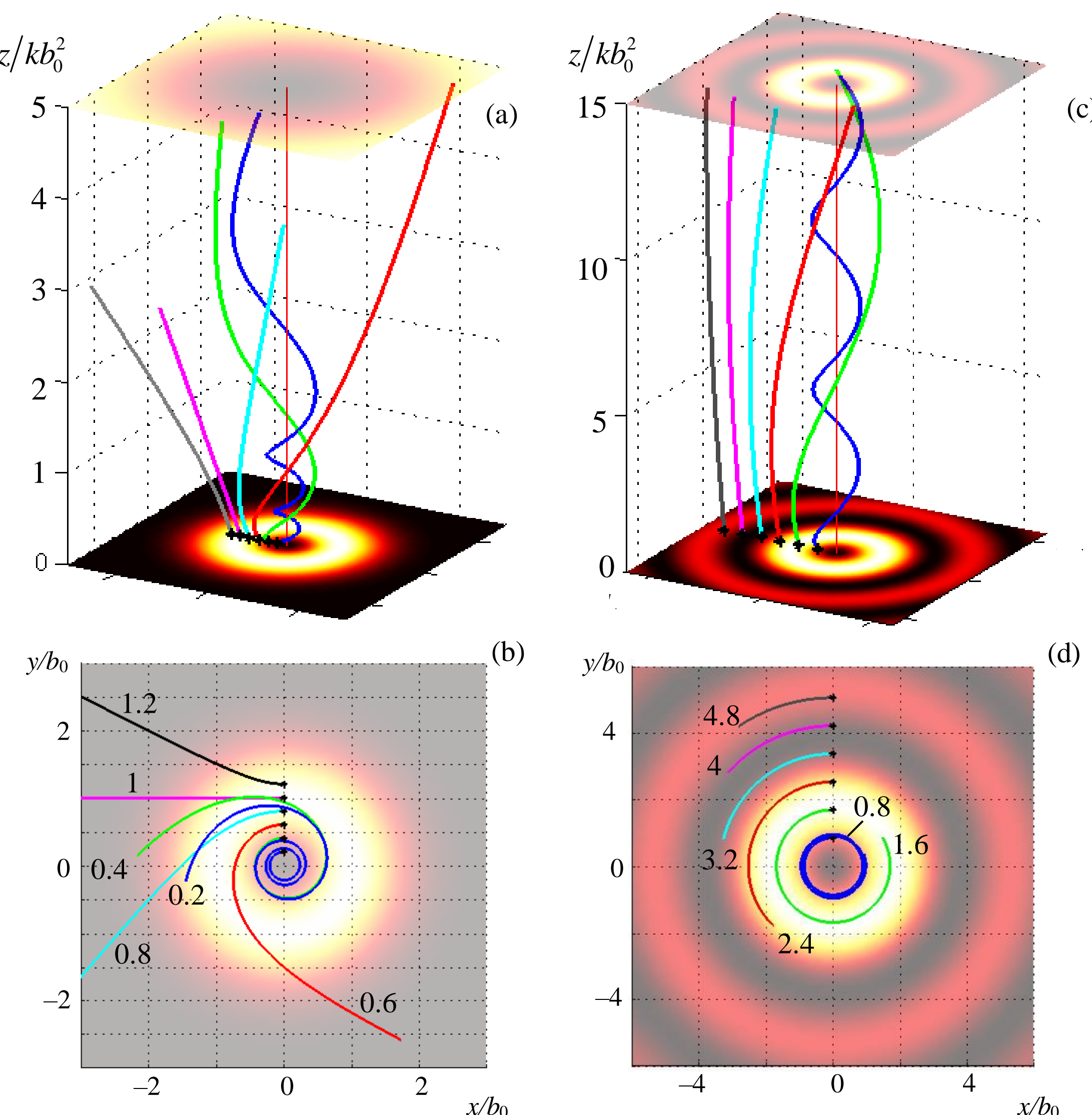

**Figure 4**. Energy flow lines in the $LG_{01}$ beam (4.2) for $q = 0$, $l = 1$ (left column) and in the Bessel beam (4.6) with $l = 1$ (right column): (a), (c) in the axonometric projection, (b), (d) as projected onto the transverse $(x,y)$ plane. Lines are labelled by initial distances from the beam axis in units of $b_0$. The background semitransparent images show the corresponding intensity distributions in the initial plane $z = 0$ (waist) and in the final planes $z = z_{max}$.

Hence, the vortex beam propagation can be treated as the vortex motion of a fluid ‘body’ whose density is determined by the mass equivalent of the electromagnetic beam energy [92]. It is worth noting that the transverse velocity distribution of this motion is quite similar to the velocity field in the vicinity of a straight-line vortex filament in fluid or to the magnetic field distribution near a straight-line electric current [96]. Among other things, this observation reveals the deep analogy between vortex motions of different nature and serves an additional evidence for the universal character of physical laws.

*4.2. Spin and orbital flows in circular vector beams*

The previous examples were related to the orbital flow in scalar beams; now we consider a combined manifestation of the both structural flow constituents in paraxial vector beams [97]. Since the spin flow is of strictly transverse character, the longitudinal orbital component $\mathbf{p}_z$ does not interfere with it, and here we consider only the transverse OFD $\mathbf{p}_{O\perp}$. Accordingly, the analysis is restricted to a fixed cross-section of the beam, while the beam patterns in other sections and diffraction transformations play no role. In this section we deal with the special class of LG beams (4.2) with $q = 0$, taken at the beam waist $z = 0$. In contrast to the previous section, the beam is supposed to be circularly polarized with helicity $\sigma = \pm 1$. Then, by using (3.37) and (3.45), we find that for the fundamental Gaussian beam ($l = 0$)

$$\mathbf{p}_S = \sigma \mathbf{e}_\phi \frac{r}{kb_0^2 c^2} I_{\sigma 0} \exp\left(-\frac{r^2}{b_0^2}\right), \quad \mathbf{p}_{O\perp} = 0 \quad (l = 0), \tag{4.8}$$

and for a LG beam ($l \neq 0$)

$$\mathbf{p}_S = -\mathbf{e}_\phi \sigma \frac{1}{|l|!} \frac{I_{\sigma 0}}{c^2} \frac{1}{kb_0} \left(\frac{r}{b_0}\right)^{2|l|-1} \left(|l| - \frac{r^2}{b_0^2}\right) \exp\left(-\frac{r^2}{b_0^2}\right), \quad \mathbf{p}_{O\perp} = \mathbf{e}_\phi \frac{1}{|l|!} \frac{I_{\sigma 0}}{c^2} \frac{1}{kb_0} \left(\frac{r}{b_0}\right)^{2|l|-1} l \exp\left(-\frac{r^2}{b_0^2}\right) \tag{4.9}$$

where $I_{\sigma 0} = \Phi_{0l} / \pi b_0^2$ is the constant associated with the beam power $\Phi_{0l}$. The total transverse flow uniting both contributions of equation (4.9) can be suitably represented as

$$\mathbf{p}_\perp = \mathbf{p}_S + \mathbf{p}_{O\perp} = \left(1 - \sigma \frac{|l|}{l} + \frac{\sigma}{l} \frac{r^2}{b_0^2}\right) \mathbf{p}_{O\perp} \quad (l \neq 0). \tag{4.10}$$

For the Gaussian beam ($l = 0$) the orbital flow vanishes but the macroscopic spin flow appears: in agreement with section 3.3 it possesses a circulatory character near the beam intensity maximum which in the considered case of a pure circular polarization coincides with the maximum or minimum of the $s_3(\mathbf{r})$ distribution (figure 5). In figure 6a, absolute value of the SFD (4.8) (curve labelled $p_S = p$) reaches its maximum when the intensity (curve $I$) shows the fastest fall-off (at $r = b_0$). In the LG beams (formula (4.9)) the picture becomes more interesting. As figures 6c, d show, in contrast to the SAM and OAM densities which usually coincide [92] with the transverse intensity distribution of circularly polarized LG beams (curves $I$), the corresponding transverse energy flows (curves $p_S$ and $p_O$) behave differently. At any $l$, the circulatory energy flows vanish on the axis ($r = 0$) and far from the axis ($r \to \infty$). In the intermediate region absolute values of the spin and orbital flows possess extrema. The OFD magnitude (4.9) has the maximum at

$$\frac{r}{b_0} = \sqrt{\frac{2|l| - 1}{2}} \quad (|l| > 0), \tag{4.11}$$

extremum points of the SFD satisfy the condition

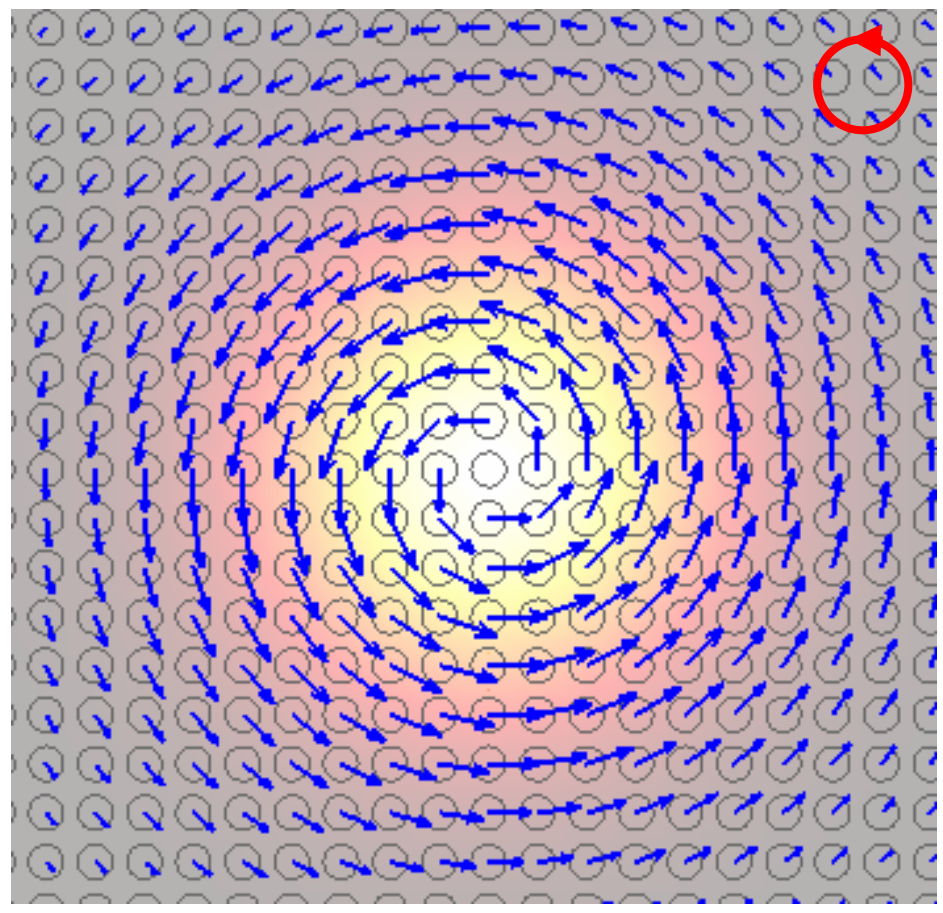

**Figure 5**. Map of the spin flow density of (4.8) for a circularly polarized Gaussian beam ($\sigma$ = 1, polarization handedness is shown in the upper right corner); lengths of arrows correspond to relative flow density, the intensity distribution and polarization ellipses (circles) are shown in the background, the beam is viewed against the propagation axis.

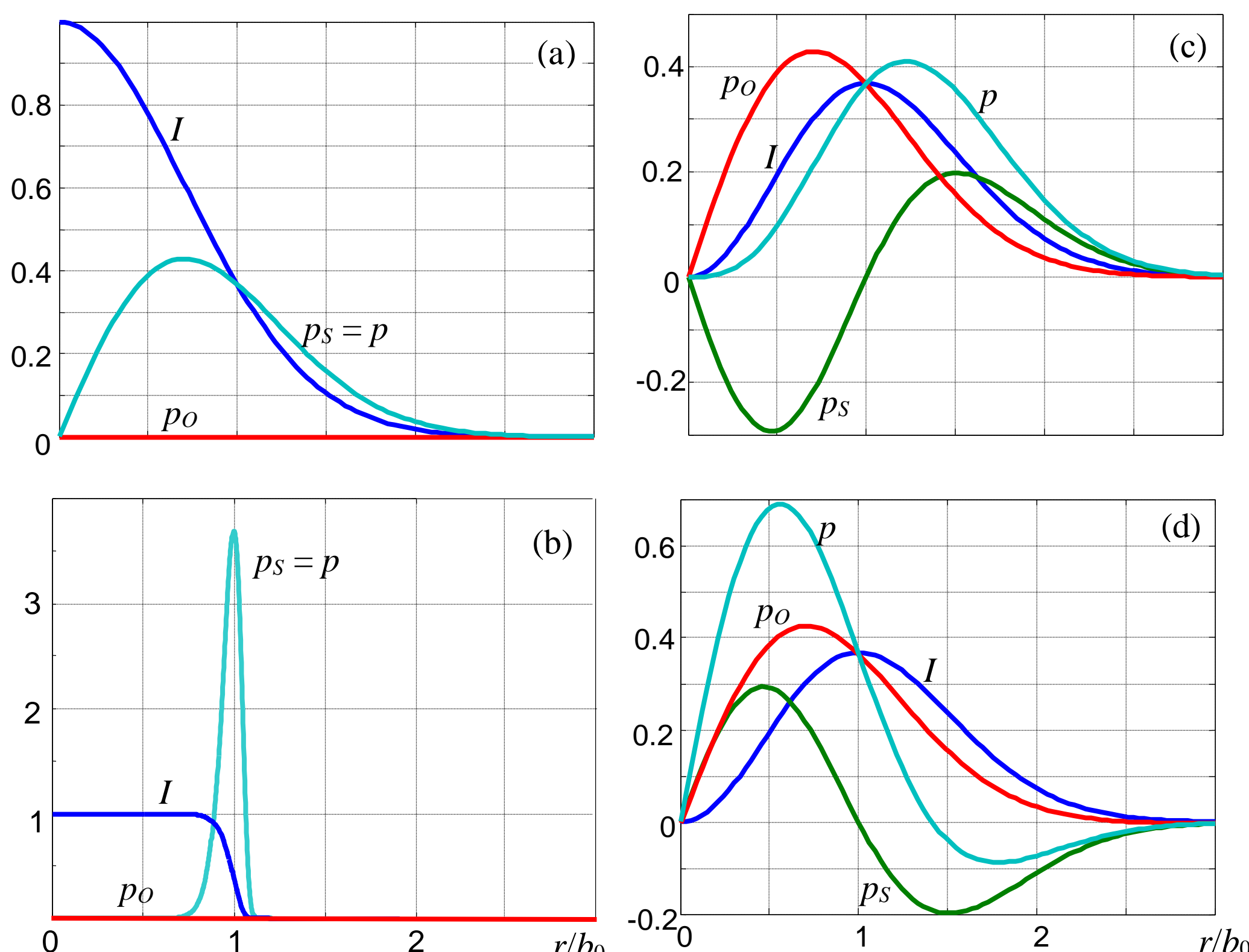

**Figure 6**. Radial profiles of ($I$) intensity (3.10) in units of $I_{\sigma 0}$, ($p_S$) SFD of first equation (4.9), ($p_O$) OFD of second equation (4.9) and ($p$) total transverse flow density (4.10) (all in units of $I_{\sigma 0}/kb_0c^2$), for the waist cross sections of circularly polarized beams: (a) Gaussian beam of figure 5, $\sigma = 1$, $l = 0$; (b) super-Gaussian beam of (4.14), $n = 20$; (c) LG beam of equation (4.2), $\sigma = 1$, $l = 1$, $q = 0$; (d) LG beam, $\sigma = -1$, $l = 1$, $q = 0$.

$$\left(\frac{r}{b_0}\right)^2 = |l| + \frac{1}{4} \pm \frac{\sqrt{16|l|+1}}{4} \qquad (4.12)$$

which corresponds to maximum gradients of the beam intensity on the inner and outer sides of the bright ring of the "doughnut" mode pattern. The expectable zero spin flow takes place at the "brightest" line of the ring.

The spin and orbital contributions may support as well as suppress each other (see figure 6c, d). In the region $r/b_0 < l$, the most important physically because it contains prevailing part of the beam power, the orbital flow dominates; otherwise (at the beam periphery) the spin contribution is more intensive. An interesting situation occurs in the near-axis region $r/b_0 << 1$ where, due to (4.9), absolute magnitudes of the SFD and OFD are almost identical. Then, if signs of $l$ and $\sigma$ coincide (that is, handedness of the macroscopic optical vortex of the LG beam and handedness of the circular polarization are the same), the total transverse energy circulation is zero at small $r << b_0$ (see figure 6c). That the spin flow can be directed oppositely to the polarization handedness, seems, at first sight, counter-intuitive but can be simply explained by the "cell model" of the spin flow formation (see section 3.3, figure 2). On the contrary, if the polarization handedness is opposite to the orbital circulation, the spin and orbital flows add constructively and provide the maximum local values of the total rotational energy flow available for circularly polarized LG beams with given $l$, as is seen from figure 6d, curve $p$.

For comparison, in figure 6b the situation of a super-Gaussian beam is presented where the intensity distribution at $z = 0$ is given by

$$I(r) = I_0 \exp\left[ -\left( \frac{r}{b_0} \right)^n \right]. \tag{4.13}$$

At $n \to \infty$ this can be a model of a sharply apertured transversely limited beam whose intensity is $I_0$ within the circle of radius $b_0$ and vanishes outside it:

$$I(r) = \begin{cases} I_0, & r \le b_0; \\ 0, & r > b_0. \end{cases} \tag{4.14}$$

The shape of this beam evolves in a rather complicated manner upon propagation along the $z$-axis [98,99], but this is not important for the present consideration within the $z = 0$ plane. The condition $n = 20$ accepted in figure 6b represents the transition from a smooth to the abrupt-boundary beam of (4.14). Like in case of figure 6a, the whole energy flow is of the spin nature but now it is concentrated in the narrow annulus near the beam boundary; this illustrates the mechanism of formation of the boundary flow (3.40).

The flow maps presented in figure 7 are in full agreement with the data of figure 6d. For considered beams, the orbital flow density possesses the same handedness in the whole cross section (compare figure 7a and curve $p_O$); however, the spin and the total flows may reverse. Note that in calculation of the full SAM over the whole cross section (e.g., by first formula (3.16)), the "opposite" spin flow of the near-axis region is compensated by the periphery contribution where the spin flow reverses. As a result, the handedness of the total SAM of the considered uniformly polarized beam always coincides with $\sigma$, which is seen from the second equation (3.16) where this compensation is ensured automatically.

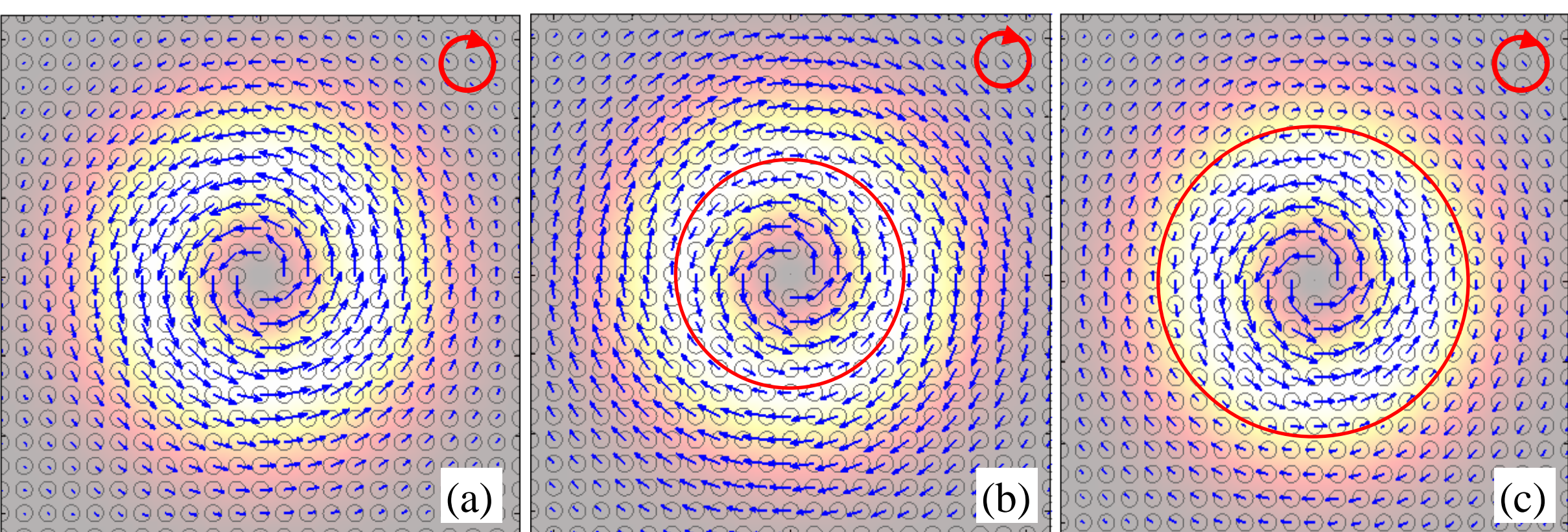

**Figure 7**. Maps of the (a) orbital $\mathbf{p}_O$, (b) spin $\mathbf{p}_S$ and (c) total $\mathbf{p}$ transverse energy flows in the cross section of a circularly polarized LG beam with $l = 1$, $\sigma = -1$ (case of figure 6b). In every point, polarization is the same as shown in the upper right corners; circular contours in panels (b) and (c) are contours where the corresponding flow component vanishes. The intensity distribution of the $LG_{01}$ mode and the polarization ellipse map are shown as a background (semitransparent).

### *4.3. Transverse optical vortex*

The next example [100–102] is again associated with the scalar model of the optical field and represents a certain generalization of the early works treating the light vorticity near the Airy rings [9,10]. The field is formed by two paraxial Gaussian beams (a special case of (4.2) with $q = l = 0$) with the common waist plane at $z = 0$ but different waist sizes $b_{10}$, $b_{20}$ and amplitudes $E_1$, $E_2$. Initial phases of the beams in the waist cross section differ by $\pi$ so that the resulting field is described by equations

$$E(\mathbf{r}, z) = E_1 \exp\left(-\frac{r^2}{2b_1^2}\right)\exp\left(i\varphi_1\right) - E_2 \exp\left(-\frac{r^2}{2b_2^2}\right)\exp\left(i\varphi_2\right), \tag{4.15}$$

$$\varphi_{1,2} = k\frac{r^2}{2R_{1,2}} - \chi_{1,2};$$

$b_{1,2}$, $R_{1,2}$ and $\chi_{1,2}$ are determined by (4.3) and (4.4). The field (4.15) is characterized by the ring of zero amplitude in the waist cross section with radius

$$r_0 = \sqrt{\frac{2b_{10}^2 b_{20}^2}{b_{20}^2 - b_{10}^2}\ln\left(\frac{E_1}{E_2}\right)}$$

(see figure 8a). On crossing this ring, the phase of the resulting field (4.15) $\varphi = \arctan\left(\operatorname{Im} E/\operatorname{Re} E\right)$ experiences a $\pi$-jump [2,100] forming the 'edge phase dislocation' [2,13]. In the dislocation vicinity, wavefronts bend (figure 8b) and the OFD lines, directed along their normals, form a sort of circulation. The separatrix loop is a boundary of the circulation area; outside it the energy flow is directed in ordinary manner. In the waist plane, the circulation area is limited by the saddle point where the macroscopic energy flow vanishes, as well as in the dislocation point. Of course, in the real 3D space, the separatrix is a toroidal surface and the saddle points form a closed ring in the waist plane. Between the dislocation ring and the saddle-point ring the light energy seem to propagate backward – this is an example of "negative propagation" which occurs also in other situations where the transverse zero-amplitude lines emerge. In particular, similar behaviour takes place near the dark rings of the Bessel beams [89] and evanescent Bessel beams formed in sandwich structures [103], near the Airy rings in the focal plane of a diaphragmed lens [10] or in front of the edge of a half-plane reflecting screen [11]. According to [100], the dimension of the circulation torus is about 0.01λ along the transverse coordinate and 0.1λ in the longitudinal direction[1].

The analogous analysis was applied to combinations of the 2D Gaussian beams where the ring edge dislocation degenerates into two straight-line dislocations [101,102]. The details of the phase surface geometry appear to be a bit different but the energy flow pattern in the plane orthogonal to the dislocation lines is qualitatively similar to the picture of figure 8b. Characteristic vortex and saddle-point structures in the Poynting vector distribution were revealed in the electromagnetic field patterns formed near sharp edges and sub-wavelength slits in the conducting plates [104–106] as well as in some model examples of 2D radiation fields [89].

---

[1] There are some precautions concerning the physical interpretation of such small-scale vortex structures, see below in section 7.1.

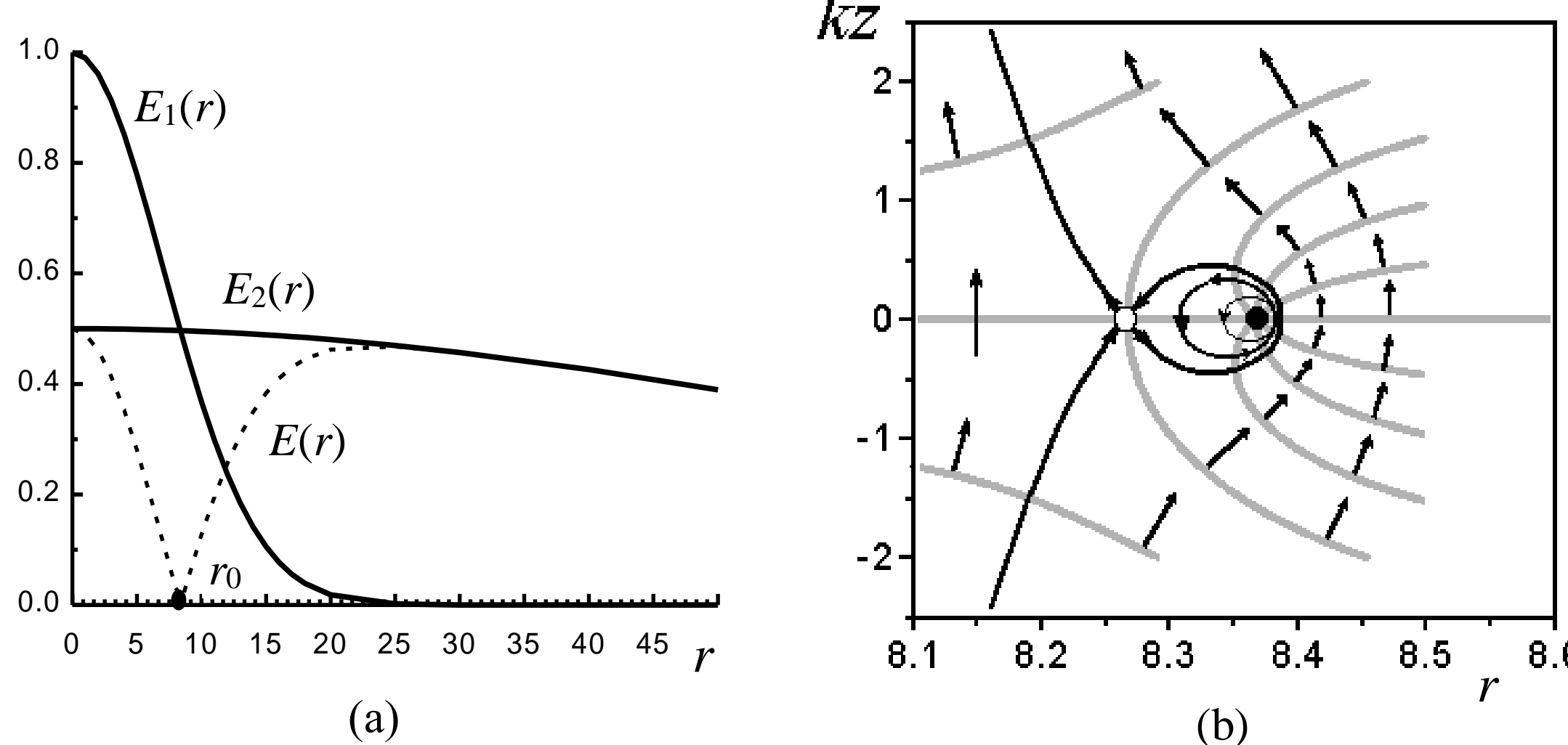

**Figure 8**. (a) Amplitude distribution of the interfering Gaussian beams $E_1(r)$ и $E_2(r)$, and the resulting amplitude $E(r)$ in the waist cross section ($z$ = 0). On the ring $r = r_0$ amplitude $E(r)$ vanishes. The beam parameters: $E_1 = 1$, $E_2 = 0.5$, $b_{10} = 7$, $b_{20} = 70$ (in units of $\lambda/2\pi$); (b) Family of equiphase lines with step $\pi/4$ and wave normals (indicated by arrows) in the edge dislocation vicinity. Point of zero amplitude is shown by filled circle, equiphase lines converge and break up there. The branching point of the equiphase line $\varphi = 0$ (saddle point) is shown by empty circle. The separatrix forms a closed ring embracing the point of zero amplitude and intersects the saddle point.

## 5. Singularities of the internal flows

A few examples considered in the previous sections show some characteristic features of the internal energy flows associated with a single polarization component of the optical field. However, figures 1 and 3–6 present rather simple artificial situations while the real (‘generic’ [13]) patterns of the energy flows are generally more complex. Their analysis and classification require the topological concepts to be employed. One should address some special points where the field pattern shows qualitative distinctions – ‘singular points’ of the internal flows. A singular point is interesting not only ‘per se’ – the matter is that it ‘organizes’ the whole field in the neighbouring space [13,107]. Besides, the singular points of various types adjoin, alternate and combine in compliance with the distinct rules [13,107–109], and knowledge of the singular points’ types and positions (so called ‘singular skeleton’ of the field) can provide essential information on the whole field, in many cases sufficient for applications. We have remarked in the Introduction that the studies of singularities in the wave structures of light beams [2,5,13,110] were among the motives for the deeper interest in the internal flow patterns; in turn, the “own” singularities of the Poynting vector fields also attract the considerable attention [87,109,111–115].

### *5.1. Singularities in 3D flows*

In accord with the common physical sense, a singular point of a function is a point where the function is indeterminate or discontinuous. The Poynting vector distribution as well as all its structural constituents $\mathbf{p}_\perp$, $\mathbf{p}_O$, $\mathbf{p}_S$ and partial contributions $\mathbf{p}_\sigma$ are regular everywhere since they are composed of the regular solutions of the Maxwell equations **E** and **H**. In this view, not the energy flow vectors themselves but some their characteristics can be singular: namely, a vector direction becomes indeterminate in points where the vector length equals to zero [17,21,89]. The singularity character is specified by possible patterns of the flow lines in its vicinity and obeys only topological regulations, which are the same for any vector field. Its description is mathematically equivalent to the theory of stability of dynamical systems [116] applied to (4.1).

In general, singular points of the Poynting vector fields can be caused by vanishing of a certain field vector (electric- or magnetic-field-induced singularities) and by the special polarization providing that $\mathbf{E}^* \times \mathbf{H} = 0$ in (2.3) and (2.5) (polarization-induced singularities) [89]. The singular points can be isolated or form 1D (line) or 2D (surface) manifolds. The issue of isolated singular points of the total momentum density $\mathbf{p}(\mathbf{r},z)$ seem to be studied insufficiently, at least for fields in homogeneous media without charges and currents [89]. The only thing one can tell definitely is that due to (2.11) the flow lines of **p** are everywhere continuous, i.e. no sources nor sinks of the 3D field **p** can exist (see the classification of 2D singular points in Table 1 below). Additionally, a number of analytical studies and numerical examples (including those presented in this paper) lead to suspicion that isolated zeros of **p** occur rather rarely and do not play a significant role in the beam propagation. Situations where the singular points "condense" in a connected set are more realistic. For example, in the field of vector Bessel beams [89,111] the singular cylindrical surfaces coaxial with the beam may exist, and on the opposite sides of these surfaces the energy moves along the opposite spirals. In the evanescent waves, a singular manifold can embrace the whole beam-occupied space [111], which is natural as there is no running wave in such beams.

In general, the detailed study of the 3D flows and their singularities is now at the early stage. As future trends, possible existence of limit cycles and stochastic dynamic regimes in the flow patterns is anticipated [89].

*5.2. Singularities in 2D flows*

In the most common situations the flow singularities form 'singular lines'; one can easily see that it is just the case for the orbital flow in the LG beams (section 4.1) where singular line coincides with the axis $z$ as well as for the transverse vortex (section 4.3) where singular lines were represented by the dislocation and the saddle-point rings. In such situations the typical near-singularity flow patterns are formed in planes orthogonal to the singular lines where the singular line degenerates into a single point. This is especially applicable to paraxial beams where singular lines, as a rule, are sorts of threads strongly stretched in the longitudinal direction [117]. For these reasons, consideration of the 2D singularities of the flow patterns deserves the special interest [21,22,112,118–121]; besides, the 2D analysis is directly applicable to the transverse flow characteristics that are so important in the study of paraxial beams (section 3).

The topological analysis of singular points is quite elementary and common for any 2D vector field; we perform the singular points' classification with the help of 2D version of the dynamic equation (4.1). Let a singular point be located, say, at $x = x_s$, $y = y_s$; generically, in its nearest vicinity flow densities can be presented in the form

$$p_x = g_{11}(x - x_s) + g_{12}(y - y_s), \quad p_y = g_{21}(x - x_s) + g_{22}(y - y_s)$$

(the flow density component indices $S$, $O$, $\sigma$, etc. are omitted for simplicity, $g_{mn}$ ($m$, $n$ = 1, 2) are real numbers). Flow lines are determined by differential equation $dy/dx = p_y/p_x$, that is

$$\frac{dy}{dx} = \frac{g_{21}(x - x_s) + g_{22}(y - y_s)}{g_{11}(x - x_s) + g_{12}(y - y_s)}.$$

Following the known theory [116], the singular point character is determined by the eigenvalues $\lambda_1$ and $\lambda_2$ of the stability matrix

$$\mathsf{G} = \begin{pmatrix} g_{11} & g_{12} \\ g_{21} & g_{22} \end{pmatrix}. \tag{5.1}$$

This matrix provides also exhaustive classification of possible topologies of flow lines in the singular point vicinity; main results in the form adapted to singular optics are summarized in Table 1. One can see examples of the flow field patterns near the singular points in figures 1, 3–5, 7, 8. The detailed quantitative description of these patterns in application to separate structural and partial flow components can be found in numerous publications [21,89,104,105,108,114,115,118].

Table 1. Classification of generic singular points in 2D vector fields [21] (following to [116])

| Condition for eigenvalues $\lambda_1$ and $\lambda_2$ of the stability matrix (5.1) | Typical view of the flow lines | Terms and short characteristic |
|---|---|---|
| At least one of $\lambda_1$ and $\lambda_2$ equals to zero (the stability matrix is degenerate) | | No singularity (regular point) |
| $\lambda_1$ and $\lambda_2$ are real and of the same sign | | Stable (source) or unstable (sink) node (flow lines go towards or outwards the singular point) |
| $\lambda_1$ and $\lambda_2$ are real and of the opposite signs | | Hyperbolic point; saddle |
| $\lambda_1$ and $\lambda_2$ are pure imaginary | | Elliptic point; centre; circulation; vortex |
| $\lambda_1$ and $\lambda_2$ are complex conjugate with non-zero real parts | | Stable or unstable spiral point (focus). Flow lines approach to or emanate from the singular point making infinite number of rotations |

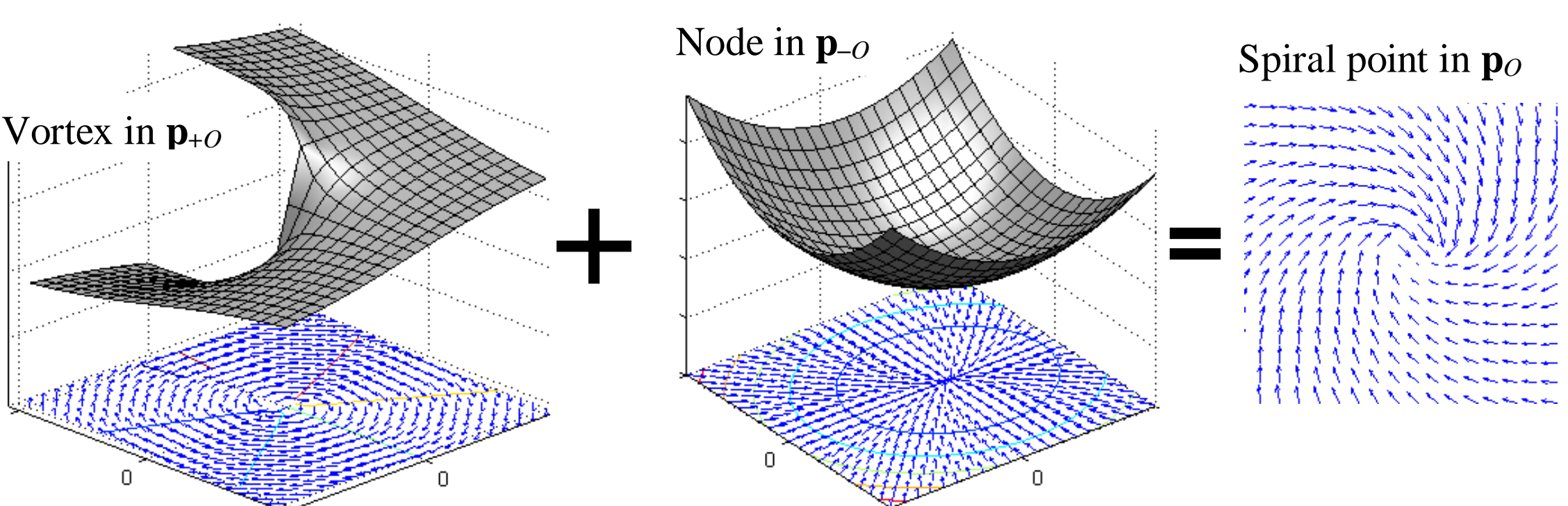

**Figure 9**. Formation of the spiral point singularity in the orbital flow pattern $\mathbf{p}_{O\perp}$ when the partial components possess closely spaced node and vortex.

In addition to the qualitative class of Table 1, each singular point is characterized by the angle $\Delta\Theta$ of the flow line rotation on the round trip near the point; the winding number $m = \Delta\Theta/2\pi$ is called the topological charge of the singularity [2,5,13]. For all the singularities of Table 1, $m = 0$, $\pm 1$, in contrast to the polarization singularities (C-points) where angle $\Theta$ characterizes the orientation of polarization ellipses and the half-integer values of $m$ are typical [13,107]. The magnitude of the transverse flow density grows linearly with the distance from the singular point, which can be represented by a conical surface [21] (a polarization-singularity analogue of this behaviour is manifested in the distribution of the polarization ellipse form-factor which is also 'conical' near a C-point and can be imaged by the 'diabolic' structures [122,123]). This conical structure can be treated as the 'domain of influence', or 'physical body' of the singularity, and the singular point 'per se' forms its 'core'. The problems of the 'singular body' dependence on the beam parameters and its evolution upon the beam propagation are fairly important for the qualitative description of complicated (in particular, stochastic) beams. To the best of our knowledge, they are marginally considered for the Poynting vector fields in the current literature. As an exclusion, one can mention the promising concept of the optical vortex morphology (parameters of anisotropy and orientation) [124–128] which is naturally applicable to the elliptic points of the 2D Poynting fields (fourth row of Table 1) and can be generalized to other singularity classes [21].

It should be noted that Table 1 represents only generic singularities [13], i.e. those that are stable and preserve the structure upon small variation of the 'free' field parameters (initial conditions). In practice, more complex singular patterns may occur, often due to combinations of "simple" singular structures belonging to separate contributions. In this process, interesting phenomena of 'collisions' and 'interactions' of singularities may affect their positions, 'class' (as specified by Table 1) and even existence. As a simple illustration, in figure 9 we present an example where the separate partial summands contain closely spaced node and vortex whose interaction contributes to the spiral point in their superposition (3.44) (the pattern details depend on the relative magnitudes of the combined components, their spatial distributions and the distance between the 'primary' singularities of the summand fields).

As follows from analyses of various models and numerical examples [17,21,22,108,112,113], 'pure' singularity classes (nodes and vortices) are more frequent in the partial fields of separate polarization components while the total transverse flow fields (and even the 'complete' SFD and OFD distributions) contain noticeable number of spiral points instead (results of interactions similar to one presented in figure 9). The saddle points seem to occur with equal frequency in the partial and total flow patterns (which is logical as the saddle points provide necessary 'links' between adjacent vortices and/or spiral points [13]). At present, we cannot derive any far-reaching conclusions from these observations, especially because both in numerical analysis and in experiments the vortices and spiral points are manifested almost identically and often cannot be distinguished reliably.

Another important question to be answered concerns relations between the energy flow singularities and the 'usual' polarization singularities (C-points of circular polarization and L-lines of linear polarization, as far as the transverse beam patterns and the transverse flows are considered [107,109]). For partial flows, such relations are quite obvious: e.g., a C-point coincides with the core of the vortex flow belonging to one of the helicity components [17,21,115] (however, addition of the flow contributed by the opposite circular polarization will destroy this vortex or, at least, move it to other position [17,21]). If the field is considered as a superposition of linearly polarized partial fields, L-lines are trajectories of the 'partial component' vortices whose positions and number depend on the chosen linear polarization basis $\mathbf{e}_x$, $\mathbf{e}_y$; they move along L-lines, annihilate and emerge when the basis unit vectors rotate [129].

In general, only positions of the field-induced flow singularities [89] coincide with the phase singularities of at least one of the orthogonally polarized field components. As a rule, singularities of the total transverse flow patterns seem to possess no simple correspondence with the C-points and L-contours [17,21]; some correlations anticipated by numerical examples [108,115] still need to be supported by further investigations.

## 6. Interaction and mutual conversion of different forms of the AM and energy flows

In section 2, the three forms of the AM of light beams were differentiated: SAM, intrinsic OAM, and extrinsic OAM, which correspond to different degrees of freedom of light and to different structural constituents of the energy flow. Different forms of AM are independent in free space and conserve upon free propagation. However, any perturbation (inhomogeneity or anisotropy of the medium, nonlinear effects, diffraction and scattering by various obstacles) can make them to be coupled – the so called spin-orbit interaction (SOI) or orbit-orbit interaction (OOI) takes place [25,33–70]. Since the SAM, intrinsic OAM and extrinsic OAM are usually associated, respectively, with the polarization, transverse profile inhomogeneity (conventionally, the 'optical vortex' structure), and trajectory of light, one can classify the SOI and OOI phenomena as mutual influence of the (i) polarization and vortex, (ii) polarization and trajectory, and (iii) vortex and trajectory. Basically, numerous manifestations of these AM interaction phenomena reduce to the facts when the spatial characteristics (profile, trajectory) of the propagating beam depend on the beam polarization state or its intrinsic OAM, and vice versa.

First of all, interactions between the SAM and OAM arise naturally in anisotropic media, where the polarization and phase (wave vector) are coupled with each other via dielectric tensor. An efficient conversion from the SAM to intrinsic OAM (i.e., from circular polarization to vortex) may appear in anisotropic media with certain azimuthal symmetries [57,64–70]. Similar interaction and spin-to-orbit conversion takes place in nonparaxial fields [23,24], upon tight focusing [51–57] and scattering by small spherical particles [48–50,56], and upon propagation in multimode optical fibers [59]. In these cases, the coupling between the polarization and wave vector originates from the transversality condition for partial plane waves forming the field. Naturally, any spin-to-orbit AM conversion implies the corresponding conversion of the energy flows. This is directly confirmed, e.g., by experiment [53], where the polarization-dependent orbital motion of probing particles in the tightly-focused circularly polarized field was observed.

Any breaking of azimuthal symmetry in nonparaxial fields [24,56,58,60–63], or in paraxial fields propagating through inhomogeneous [33–43] or anisotropic [69] media or even in free space [46,47] results in polarization- or vortex-dependent transverse shift of the field centroid (trajectory), i.e., conversion from intrinsic AM (either SAM or OAM) to the extrinsic OAM [24,33,34,36–38,41,42,46]. Such transverse shifts are called spin- or orbital-Hall effects of light. The Hall effects of light arise upon reflection or refraction of paraxial beams at dielectric interfaces [33–43], propagation of paraxial light in gradient-index media [33,34,44,45], tilt of the beam with respect to the detector or diffraction grating [40,46,47,139], and in asymmetric focusing or scattering configurations [24,56,58,60–63].

Below we consider the simplest typical examples of the AM interaction and conversion related to the intrinsic geometry of light propagation in free space rather than to particular properties of the medium (anisotropy, etc.).

### *6.1. Spin-orbit interaction in strongly focused beams*

First, we consider the interaction between the SAM and intrinsic OAM in an azimuthally symmetric non-paraxial field. Such a situation occurs, e.g., upon tight focusing of a circularly-polarized paraxial field by a high-numerical-aperture lens [9,51–56], illustrated by figure 10a. As it is known from the Richardson-Wolf theory [9], the role of the lens basically consists in meridional redirection of the wave vectors of partial plane waves (geometrical-optics rays) refracted at the lens. This transformation is purely geometric: it does not change the polarizations of a partial wave in the ray-accompanying coordinate frame [56] and reveals universal features of non-paraxial polarized fields in free space [22,24].

The transverse Cartesian components of the incident paraxial circularly-polarized electric field $\mathbf{E}_0$, which propagates along the $z$-axis, can be written as

$$E_{0x} = u/\sqrt{2}\,,\quad E_{0y} = i\sigma u/\sqrt{2} \tag{6.1}$$

where $\sigma = \pm 1$ and $u \equiv u_\sigma(\mathbf{r})$ is the corresponding complex amplitude (3.2) in the circular basis and we neglect the longitudinal component of the incident field (see equations (3.3) – (3.5)). The beam intensity is $I_\sigma^0 = cg|u|^2$. For simplicity, we assume that $u(\mathbf{r})$ does not depend on azimuthal coordinate $\phi$, and, hence, the circulatory part of the orbital energy flow (the first summand of (3.45)) and the OAM vanish for the incident paraxial field:

$$\mathbf{p}_{0O}\cdot\mathbf{e}_\phi = J_{0O} = 0\,. \tag{6.2}$$

The intrinsic spin flow is determined by (3.37), $\mathbf{p}_S^0 = -\dfrac{g\sigma}{2\omega}\mathbf{e}_\phi\dfrac{\partial|u|^2}{\partial r}$, which yields the SAM value per unit $z$-length in agreement with (3.16) and (3.42), (3.43)

$$J_{0S} = \frac{\sigma}{\omega}W\,,\quad W = 2\pi g\int_0^\infty |u|^2\, r dr\,. \tag{6.3}$$

To describe the focused field after the lens, we employ the simplified model of [55] together with geometric arguments of [56] illustrated by figure 10. We start with transition to the radial and azimuthal components of the field (6.1):

$$E_{0r} = ue^{i\sigma\phi}/\sqrt{2}\,,\quad E_{0\phi} = i\sigma ue^{i\sigma\phi}/\sqrt{2}\,. \tag{6.4}$$

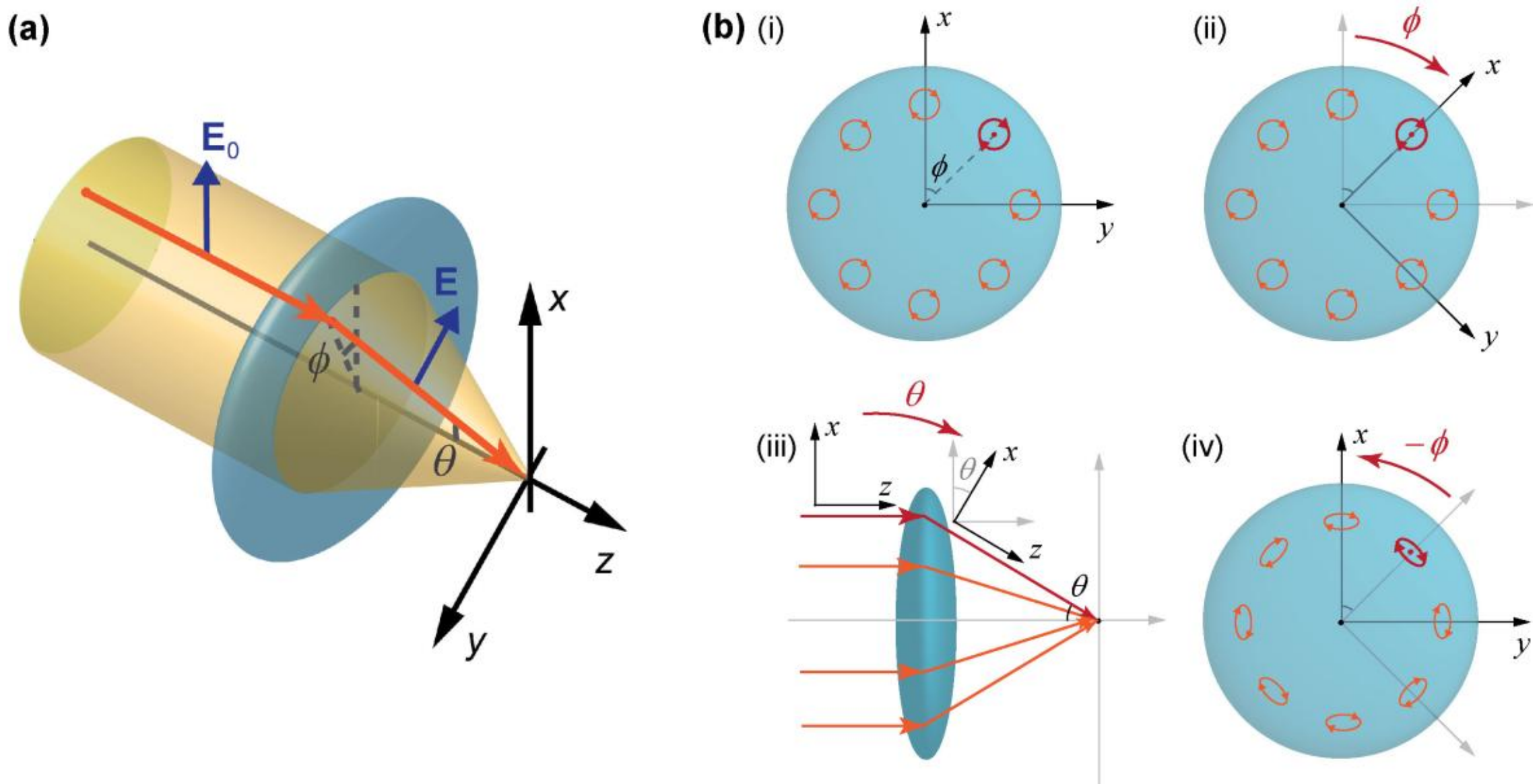

**Figure 10.** Transformations of the electric field of a circularly-polarized wave focused by a lens [56]. (a) Geometry of the problem. (b) Successive rotations of the local coordinate frame describing the polarization transformation upon focusing: Azimuthal rotation (i)–(ii) corresponds to transition to the radially oriented local frame, polar rotation (iii) describes refraction of the partial rays by the lens, back azimuthal rotation (iv) signifies transition to the original global coordinate frame. The resulting transverse polarization distribution represents azimuthally oriented ellipses, whereas the longitudinal field component exhibits charge-1optical vortex.

Note that these components possess azimuthal helical phases $\sigma\phi$ caused by the rotation of the local coordinate frame with radially-oriented $x$-axis (see figure 10b). The focusing re-directs the radial component of the field (6.4) leaving the azimuthal one invariant, so that

$$E_{0r} \to (E_r, E_z) = (E_{0r}\cos\theta,\, E_{0r}\sin\theta)\,. \tag{6.5}$$

Here the focusing (refraction) angle $\theta$ is connected with the radial coordinate $r$: $\theta = \tan^{-1}(r/f)$.

After transforming the components of the focused field (6.5) back to the global Cartesian frame, the field components after the lens can be written in the form

$$E_x = ue^{i\sigma\phi}\left(\cos\theta\cos\phi - i\sigma\sin\phi\right)/\sqrt{2}\,,$$
$$E_y = ue^{i\sigma\phi}\left(\cos\theta\sin\phi + i\sigma\cos\phi\right)/\sqrt{2}\,, \qquad (6.6)$$
$$E_z = ue^{i\sigma\phi}\sin\theta/\sqrt{2}\,.$$

These equations describe the field in the fixed cross section located immediately after the lens, which is considered as a phase screen. The propagation and interference of partial waves upon actual focusing, i.e. the field concentration in the focal spot, is described by Debye integral and is not analysed here. Indeed, the AM conversion occurs upon the field refraction at the lens and the AM balance is unchanged upon further propagation and interference in free space. Note that relations (6.6) contain the spatial inhomogeneity of the focused field even if the incident beam is homogeneous, $u(\mathbf{r}) = \text{const}$. In particular, figure 11 shows that spatial distributions of the transverse components lose the initial radial symmetry, whereas the longitudinal *z*-component carries a $\sigma$-dependent vortex.

It is helpful to write the field components (6.1) and (6.6) in the global helicity basis (3.4) $(\mathbf{e}_+, \mathbf{e}_-, \mathbf{e}_z)$. Assuming, e.g., right-hand circular polarization of the incident wave ($\sigma = +1$), the corresponding field components $(E_+, E_-, E_z)$ ($E_\pm = (E_x \mp iE_y)/\sqrt{2}$) for the fields before and after the lens take the form [56]

$$(E_{0+}, E_{0-}, E_{0z}) = (1,0,0)\,u\,, \quad (E_+, E_-, E_z) = \left(a_1, -a_2 e^{2i\phi}, -\sqrt{2a_1a_2}\,e^{i\phi}\right)u\,, \qquad (6.7)$$

where $a_1 = \cos^2(\theta/2)$ and $a_2 = \sin^2(\theta/2)$. Equation (6.7) shows that pure right-hand polarized incident field effectively acquires left-hand polarized (when projected onto the $(x, y)$-plane) component $-a_2 e^{2i\phi} u$ with the charge-2 optical vortex $e^{2i\phi}$ and the longitudinal $z$-component $-\sqrt{2a_1a_2}\,e^{i\phi}u$ with the charge-1 vortex $e^{i\phi}$ (see figure 11). We emphasize that these components arise as a result of purely geometric local transformations, and each partial plane wave does not change its polarization in the ray-accompanying coordinate frame. At the same time, as we will see, these geometrical changes of partial fields bring about real physical conversion between different forms of the energy flow and AM. Optical vortices of the left-hand polarized transverse component and longitudinal component of the focused field signify non-zero OFD circulation and non-zero OAM. It is worth remarking conservation of the sum of the SAM and OAM quantum numbers $[\sigma, l]$: $\sigma + l = \text{const}$. According to (6.7), the initial field with $[1,0]$ is partially transformed to $[-1,2]$ ($E_-$-component) and $[0,1]$ ($E_z$-component, which is always "linearly-polarized").

In general, the focusing transformation from the incident field (6.1) to the focused field (6.6) can be written in the operator form, $\mathbf{E} = \hat{T}\mathbf{E}_0$, where operator $\hat{T}$ consists of geometrical rotation transformations shown in figure 10b [56]: $\hat{T} = \hat{R}_z(-\phi)\hat{R}_y(-\theta)\hat{R}_z(\phi)$. Here $\hat{R}_a(\beta)$ is the operator of the rotation of the coordinate frame about the $a$-axis by the angle $\beta$, and the three successive rotations describe transition to the local radial-azimuthal coordinates, refraction therein, and the back transition to the global basis. In this manner, azimuthal rotations generate helical geometric phases for circularly-polarized components, whereas the meridional rotation (refraction) "squeezes" the projection of the polarization circle into an ellipse, i.e., effectively generates the opposite-helicity component. In the helicity basis, applying to the field components $(E_+, E_-, E_z)$, the transformation $\hat{T}$ acquires the following matrix form:

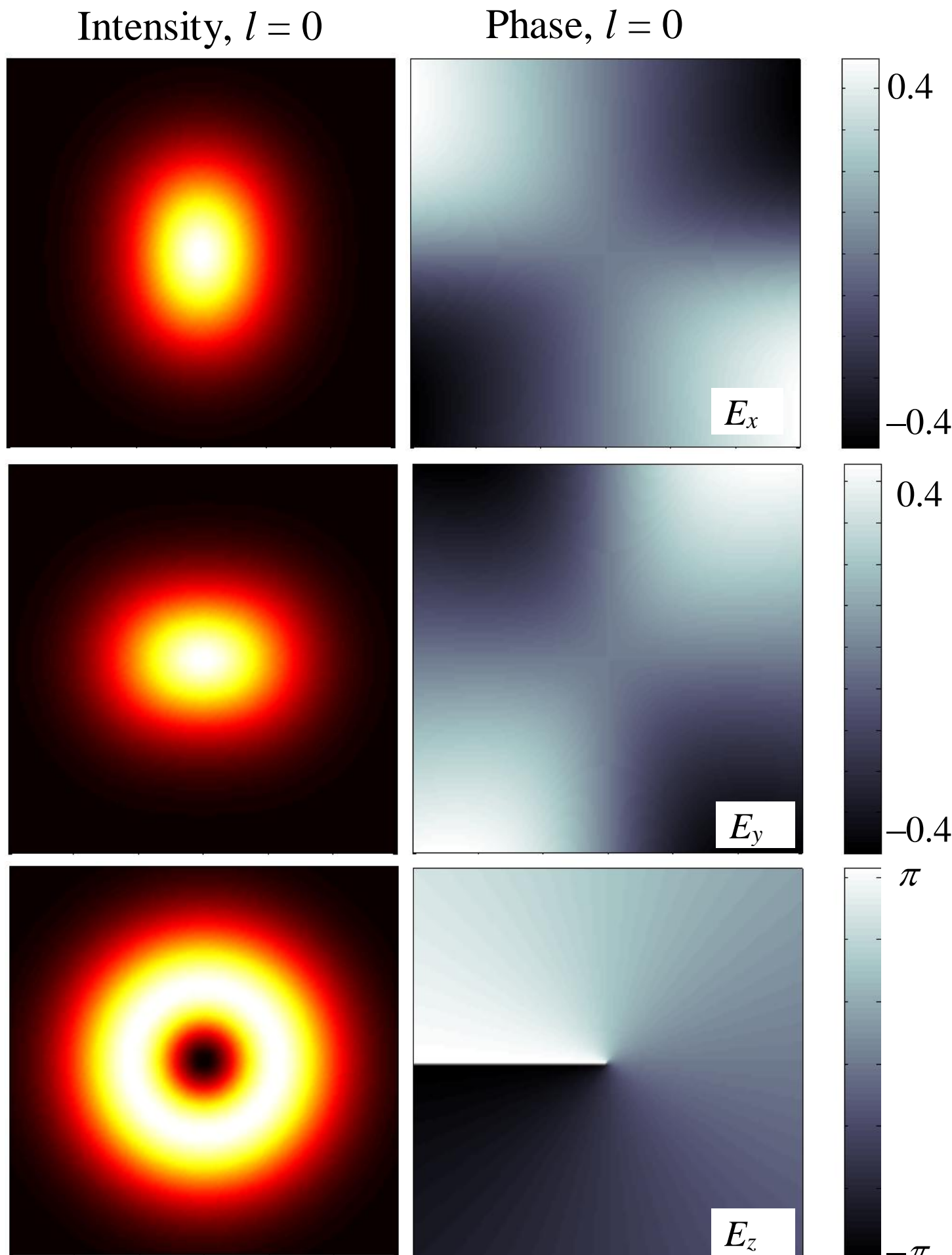

**Figure 11**. Intensity (left column) and phase (right column) profiles of the transformed right-polarized ($\sigma = +1$) Gaussian beam with initial amplitude distribution (4.2) with $q = l = 0$, calculated in accordance with (6.6) for $f/b_0 = 0.75$: (top row) component $E_x$, (middle row) component $E_y$, (bottom row) component $E_z$.

$$\hat{T} = \begin{pmatrix} a_1 & -a_2 e^{-2i\phi} & \sqrt{2a_1 a_2}\, e^{-i\phi} \\ -a_2 e^{2i\phi} & a_1 & \sqrt{2a_1 a_2}\, e^{i\phi} \\ -\sqrt{2a_1 a_2}\, e^{i\phi} & -\sqrt{2a_1 a_2}\, e^{-i\phi} & a_1 - a_2 \end{pmatrix}. \tag{6.8}$$

Here the off-diagonal elements describe transitions and conversions between different polarization components, which is accompanied by generation of helical phases producing the orbital circulatory flow and the OAM. The Jones transformation matrices of the form similar to the upper left $2\times 2$ "transverse" block of $\hat{T}$ are also typical for spin-to-orbit AM conversions in anisotropic media [64,67,68]. There, conversion occurs due to dynamical phase difference between different components of a paraxial field in anisotropic medium. In our case, effect originates from the geometric phase difference in non-paraxial field in free space; it disappears in the paraxial limit $\theta \to 0$.

Equations similar to (6.6) – (6.8) are well known in the vector diffraction theory of light focusing [9,130–133] but there they represent the Fourier amplitudes (2.17) and (2.18) of the focused field, while here we interpret them as the proper field distributions. This unveils the simplified character of our toy model being in essence of the geometrical-optics origin. The model of [55] describes only the field transformation at the lens interface and reveals the universal geometric nature of the spin-to-orbital conversion. The evolution of the wave field upon further propagation towards the focal plane is not considered here, and the issues of apodization and of the choice of the pupil function [130,132] are omitted. One can also remark that we could trace the magnetic-field transformation in the same manner and obtain the focused field representation formally identical to (6.6) but with the magnetic field components. However, both representations cannot be correct simultaneously: the magnetic field found from Maxwell equation $\mathbf{H} = -(i/k)\,\mathrm{rot}\,\mathbf{E}$ with the electric field given by (6.6) differs from equation (6.6) with replaced $E \to H$. This is an artefact of our toy model; when treating results (6.6) and (6.7) as Fourier amplitudes of the field, the 'electric – magnetic democracy' is restored [17], because the Fourier amplitudes (2.17) of the electric and magnetic fields differ by a constant phase in the helicity basis [24].

Now we proceed to calculations of the energy flows and AM of the focused field in our model. Due to its approximate character we may use the purely electric representation [17] of the AM decomposition (2.8), (2.9). The orbital flow of the non-paraxial field can be divided into contributions from the longitudinal and transverse field components:

$$\mathbf{p}_{OL} = \frac{g}{\omega}\,\mathrm{Im}\left(E_z^* \nabla_\perp E_z\right), \tag{6.9}$$

$$\mathbf{p}_{OT} = \frac{g}{\omega}\,\mathrm{Im}\left(E_x^* \nabla_\perp E_x + E_y^* \nabla_\perp E_y\right). \tag{6.10}$$

In application to the field (6.5) and (6.6) with the azimuth-independent $u(\mathbf{r})$, relations (6.9) and (6.10) yield

$$\mathbf{p}_{OT} = \mathbf{e}_\phi \frac{\sigma g}{2\omega r}|u|^2 (1-\cos\theta)^2, \quad \mathbf{p}_{OL} = \mathbf{e}_\phi \frac{\sigma g}{2\omega r}|u|^2 \sin^2\theta, \tag{6.11}$$

so that the total OFD is

$$\mathbf{p}_O = \mathbf{p}_{OT} + \mathbf{p}_{OL} = \mathbf{e}_\phi \frac{\sigma g}{\omega r}|u|^2 (1-\cos\theta). \tag{6.12}$$

The spin flow of the field is of a purely transverse character and the SFD equals to

$$\mathbf{p}_S = \sigma \frac{g}{2\omega}\left[\mathbf{e}_x \frac{\partial}{\partial y}\left(|u|^2 \cos\theta\right) - \mathbf{e}_y \frac{\partial}{\partial x}\left(|u|^2 \cos\theta\right)\right], \tag{6.13}$$

or

$$\mathbf{p}_S = \mathbf{e}_\phi \frac{\sigma g}{2\omega}\frac{\partial}{\partial r}\left(|u|^2 \cos\theta\right) = \mathbf{e}_\phi \frac{\sigma g}{2\omega}\cos\theta \left(|u|^2 \frac{r}{r^2+f^2} - \frac{\partial |u|^2}{\partial r}\right). \tag{6.14}$$

The flows (6.11), (6.12), and (6.14) are azimuthally symmetric and depend only on $r$. Their radial distributions are shown in figure 12 for the incident beam profile (4.13). Before focusing, the internal flow was of the spin nature, see figures 6a and 6b for Gaussian and flat-top paraxial beams. After refraction, i.e. exactly at the lens output, diffraction is yet unable to change the initial intensity distribution; however, the non-zero orbital flows arise in both cases (figures 12a and 12b), mostly owing to the longitudinal-component contribution. One can also notice that after focusing of the flat-top beam, the initial boundary spin flow (figure 6b) diminishes and certain volume SFD appears (figure 12b).

Substituting energy flows (6.11) and (6.14) into equations (3.15), we calculate the $z$-components of the SAM and OAM per unit $z$-length in the focused field. The results can be suitably expressed via ratios to the SAM of the incident field (6.3):

$$\Lambda_\alpha = \frac{J_\alpha}{J_{0S}} = \frac{2\pi}{J_{0S}}\int_0^\infty |\mathbf{p}_\alpha|\, r\,dr, \tag{6.15}$$

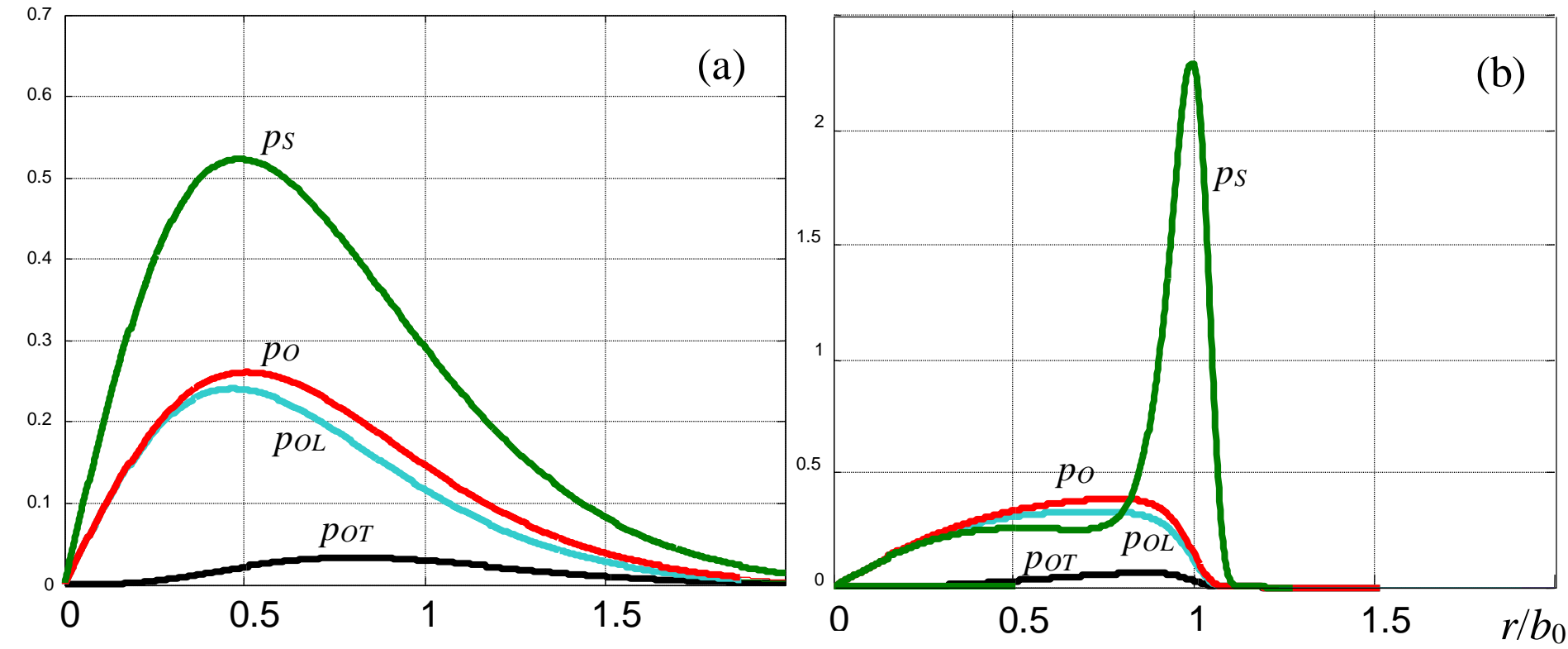

**Figure 12**. Radial distribution of partial energy flows (6.11) – (6.14) in the focused field ($f/b_0 = 0.75$, $\sin\theta_b = 0.8$) for incident beams with intensity distribution (4.13): (a) $n = 2$ (Gaussian beam); (b) $n = 20$ (approximation of the beam with abrupt boundary).

where index $\alpha$ = *OT*, *OL*, *O* or *S* regarding the respective energy-flow contributions (6.11) – (6.14). Simple analytical results can be obtained for a flat-top beam with constant amplitude within a circle of radius $b_0$ (equation (4.14)). In this case, substituting expressions (6.11) and (6.12) with (4.14) into (6.15), we arrive at

$$\Lambda_{OT} = \frac{1}{2}\left(1+\tau^2\ln\frac{1+\tau^2}{\tau^2} - 4\tau\sqrt{1+\tau^2} + 4\tau^2\right), \tag{6.16}$$

$$\Lambda_{OL} = \frac{1}{2}\left(1-\tau^2\ln\frac{1+\tau^2}{\tau^2}\right), \tag{6.17}$$

$$\Lambda_O = \Lambda_{OT} + \Lambda_{OL} = 1 - 2\tau\sqrt{1+\tau^2} + 2\tau^2 = 1 - 2\frac{\cos\theta_b\left(1-\cos\theta_b\right)}{\sin^2\theta_b}. \tag{6.18}$$

where $\tau = f/b_0 = \cot\theta_b$ and $\theta_b$ is the aperture angle.

Since the beam (4.14) possesses abrupt boundary, the SAM of the focused beam consists of two parts. The first one is associated with the volume spin flow inside the considered cross-section area,

$$\Lambda_{SV} = \tau\left(\sqrt{1+\tau^2} + \frac{\tau^2}{\sqrt{1+\tau^2}} - 2\tau\right) = \frac{\cos\theta_b\left(1-\cos\theta_b\right)^2}{\sin^2\theta_b}. \tag{6.19}$$

The second one represents the boundary contribution (3.40)

$$\Lambda_{SB} = \frac{g}{2\omega}\frac{1}{J_{0S}}\oint\limits_{b_0}\mathrm{Im}\left(\mathbf{E}^*\times\mathbf{E}\right)_z\left|\mathbf{r}\times d\mathbf{r}\right| = \frac{1}{2\pi b_0^2}\oint\limits_{b_0}\cos\theta\left|\mathbf{r}\times d\mathbf{r}\right| = \frac{\tau}{\sqrt{1+\tau^2}} = \cos\theta_b, \tag{6.20}$$

where the integral is taken along the circle bounding the beam cross section. The total SAM of the focused beam is

$$\Lambda_S = \Lambda_{SV} + \Lambda_{SB} = 2\tau\left(\sqrt{1+\tau^2} - \tau\right) = 2\frac{\cos\theta_b\left(1-\cos\theta_b\right)}{\sin^2\theta_b}. \tag{6.21}$$

The sum of the SAM (6.21) and OAM (6.18) satisfy the conservation law for the *z*-component of the total AM in the axially symmetric system:

$$\Lambda_O + \Lambda_S = 1, \text{ i.e., } J_O + J_S = J_{0S}. \tag{6.22}$$

Figure 13 shows behaviour of different AM contributions (6.16) – (6.21) in the focused field as dependent on the aperture angle $\theta_b$. Naturally, the orbital contributions vanish at $\theta_b \to 0$, whereas in the opposite limit $\theta_b \to \pi/2$ the SAM is fully converted to OAM, cf. [24,57]. In the latter case, the OAM is equally distributed between the transverse-field and longitudinal-field partial contributions. Numerical calculations for a Gaussian incident beam predict quite similar behaviour, with the only exception that the boundary term $\Lambda_{SB}$ vanishes, and the spin flow has purely volume character.

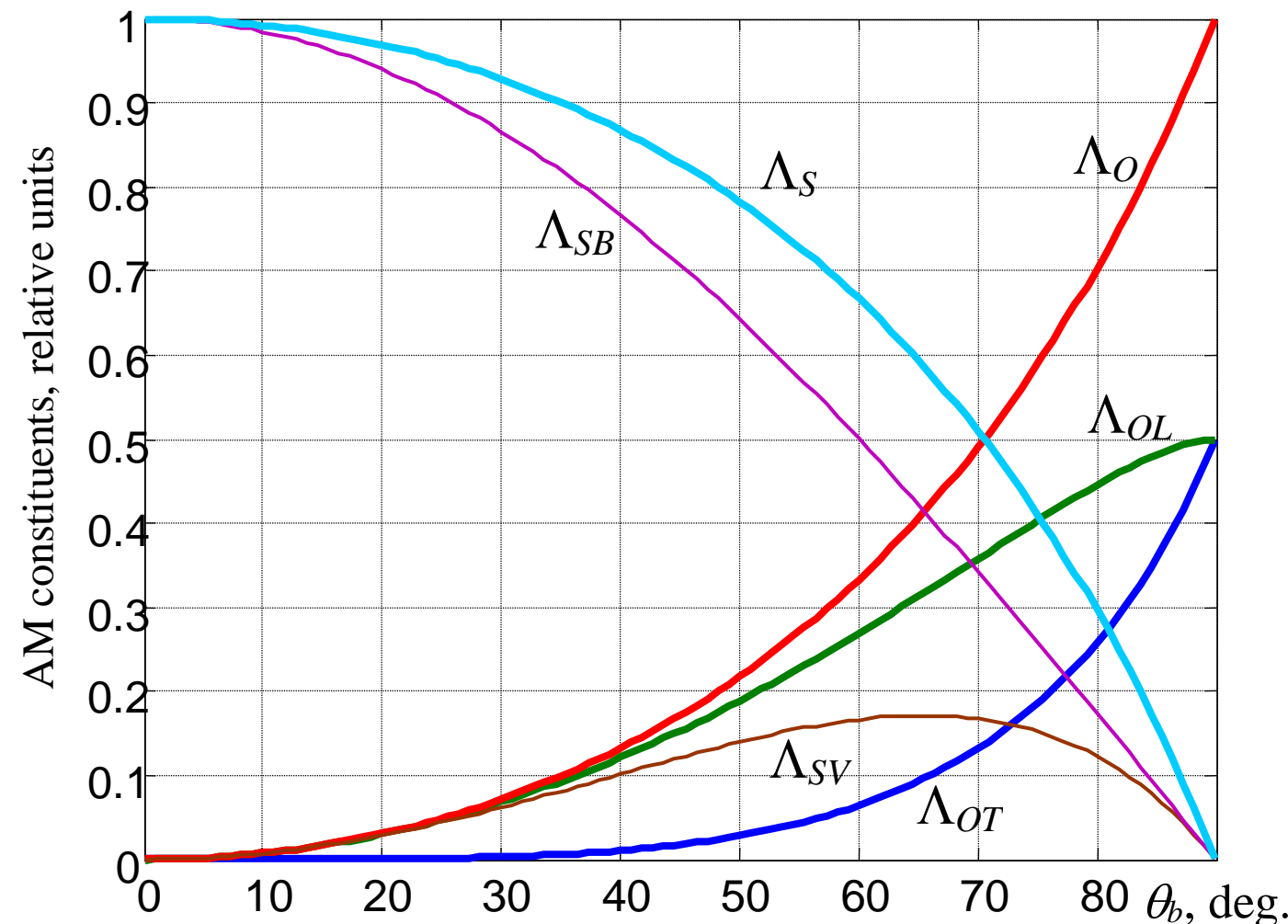

**Figure 13**. AM constituents of the focused spatially homogeneous circularly polarized beam vs aperture angle; each curve is marked by the corresponding quantity notation from (6.16) – (6.20).

Results (6.16) – (6.22) and figure 13 describe the AMs of the field just after the lens but, since the SAM and OAM conserve upon free propagation [54], they are valid in the whole space after the lens, including the focal point, most interesting for experiments [53,56,57]. More realistic models where the field is calculated by the vector Debye-Wolf integral [51,53,56] or by multipole expansion of the focused field [54,134,135] confirm the dependencies illustrated by figure 13. An advantage of the toy model of [55] presented in this section is that it gives easy access to the pattern of internal flows and immediately shows mutual conversion of their spin and orbital structural constituents.

If the incident circularly polarized beam already possesses non-zero OAM, it adds algebraically to the result of conversion, and, after focusing, the incident OAM can be amplified or weakened [52,53,55].

### *6.2. Transverse AM and Hall effects associated with oblique sections of light beams*

As it was mentioned in the beginning of section 6, any breakdown of the initial axial symmetry of a beam carrying AM can generate an AM-related shift of the beam centroid – the Hall effect of light. In particular, breaking the symmetry in nonparaxial focusing (or scattering) systems considered in the previous sub-section, one can observe the transverse shift of the focal spot, which is proportional to the AM of the incident field [24,56,58,60–63]. This effect is essentially connected with generation of the extrinsic OAM (2.28) and (3.25) [24,33,34,36–38,41,42,46]. The Hall effects arise already in paraxial fields. Perhaps, the simplest example of the Hall effect in the paraxial field, which we consider below, occurs upon observation of an oblique cross section of a paraxial beam [46,47]. This situation is typical for all problems where the detector or scattering interface has the normal tilted with respect to the beams axis, e.g., upon reflection or refraction of light at a dielectric plane interface [33–43].

Let us consider an axially-symmetric paraxial beam carrying the intrinsic AM (either SAM or OAM) $\mathbf{J} = J\mathbf{e}_z$ along its own axis $z$. Now let a detector be located in the plane $(x', y')$ with the normal $z'$-axis tilted with respect to $z$, figure 14. Naturally, the transverse AM components $J_{x'}$ and $J_{y'}$ emerge in the $(x', y', z')$ coordinate frame: $\mathbf{J} = J_{x'}\mathbf{e}_{x'} + J_{y'}\mathbf{e}_{y'} + J_{z'}\mathbf{e}_{z'}$. Remarkably, these transverse AM components turn out to be of extrinsic rather than intrinsic nature and are related to the shift of the beam centroid in a tilted reference frame [46]. The beam position can be characterized via the centroid of the energy flow across the detector, which is given by $p_{z'}(\mathbf{r}', z')$ ($\mathbf{r}' = (x', y')$):

$$\mathbf{r}'_c(z') = \frac{\int \mathbf{r}' p_{z'}(\mathbf{r}', z') d^2\mathbf{r}'}{\int p_{z'}(\mathbf{r}', z') d^2\mathbf{r}'}. \tag{6.23}$$

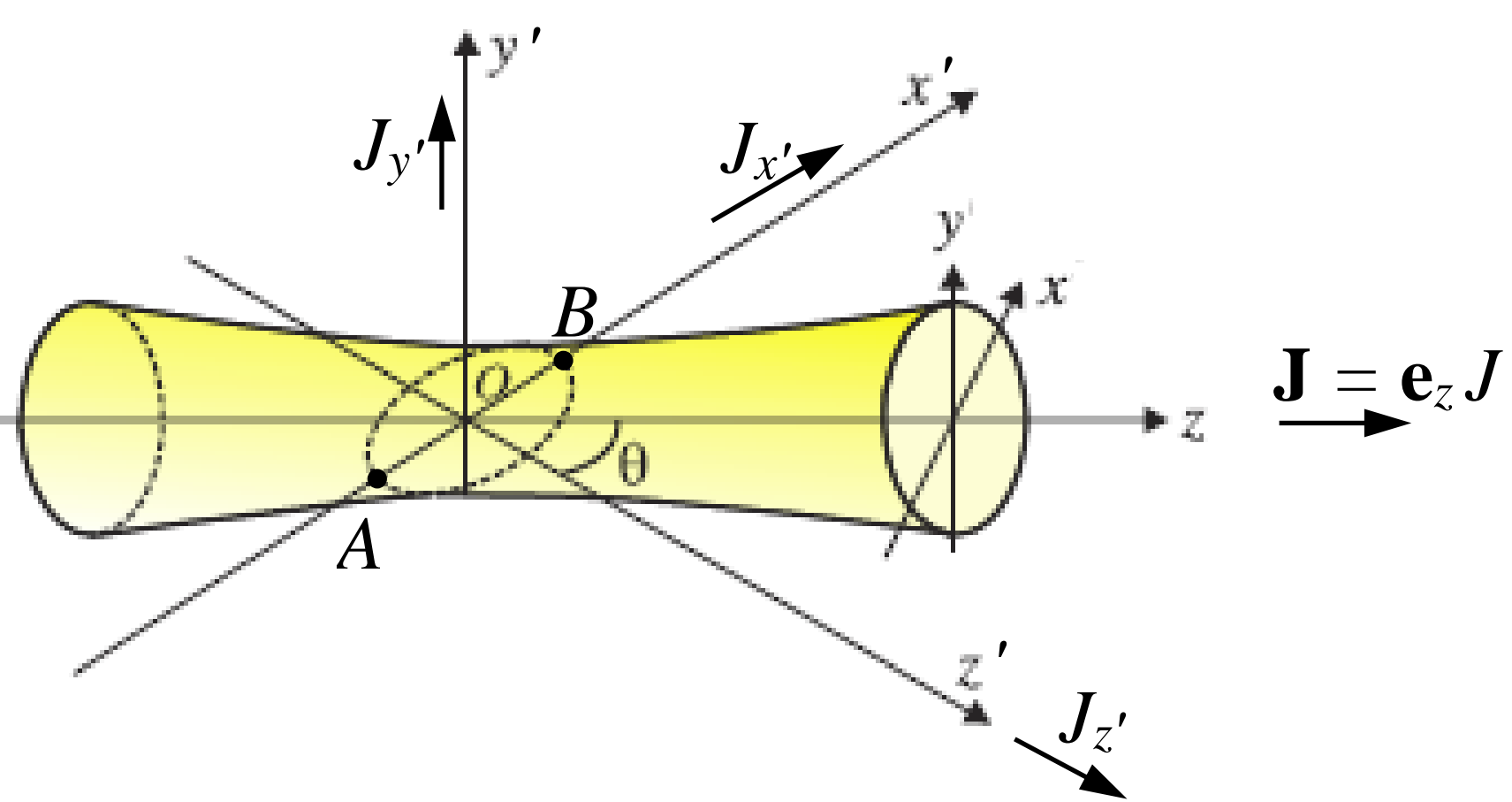

**Figure 14**. Geometrical scheme of the oblique cross-section of a paraxial beam carrying AM [46] (explanations in text).

It should be remarked that the definition (6.23) differs, in the general case, from the definition of the centroid (3.24) given in terms of the energy density $w$. They coincide only in the paraxial propagation when, in accordance to (3.9), $p_{z'} = p_z \propto w$ (see discussion in [46]). Using expressions (2.23) and (2.24) for the beam momentum and AM, the centroid (6.23) can be written as [46]

$$\Delta x'_c = -\frac{J_{y'}}{P_{z'}}, \quad \Delta y'_c = \frac{J_{x'}}{P_{z'}}. \tag{6.24}$$

Thus, in the tilted reference frame the centre of the beam is shifted in the transverse direction, orthogonal to the plane of the tilt.

For instance, let us take a circularly polarized paraxial beam bearing the momentum (3.12) $\mathbf{P} = \mathbf{e}_z W/c$ and the SAM (6.3) $\mathbf{J}_S = \mathbf{e}_z \sigma W/\omega$. Without loss of generality we assume that $y = y'$ and the $z'$-axis lies in the $(x, z)$ plane making the angle $\theta$ with the beam $z$-axis (figure 14), so that $P_{z'} = W\cos\theta/c$ and $J_{x'} = \sigma W \sin\theta/\omega$. Then, according to (6.24)

$$\Delta x'_c = 0, \quad \Delta y'_c = \frac{\sigma}{2k}\tan\theta \tag{6.25}$$

(the multiplier 1/2 appears due to consistent account for the Gaussian beam inhomogeneity [46]). This small transverse shift is of the order of a fraction of the wavelength, which is typical for the Hall effect of light. Nonetheless, such fine effects are very important for the modern nano-optics, they carry subwavelength information about light, and can be significantly enhanced using various techniques [39,44,56]. As all the SOI effects in isotropic media, the transverse shift (6.25) has universal geometric character and takes place not only in optical beams but also, e.g., in atomic or particle beams.

The experimental observability of the shift (6.25) relies on the detector's ability to register the light momentum component normal to the detector plane. This is a rather special property; in the optical domain, most of detectors are sensitive to the energy density (with polarization filters, the partial energy density of a certain polarization component can be measured). Another important assumption implied in the above analysis is that a detector located in the plane $z' = 0$ does not perturb the field of the beam propagating along the $z$-axis. In actual fact, for a freely propagating beam, the optical field in point, say, $B$ (figure 14) with $z = z_B$ is formed by contributions of all the precedent transverse sections with $z < z_B$. Then, the field that reaches the detector and interacts with it (e.g., absorbed by it) 'earlier' (say, at point $A$ with $z = z_A$ in figure 14), plays no role (or, at least, participates less efficiently) in forming the field in points with $z > z_A$. Hence, in reality, the field pattern measured in the oblique plane is rather a result of a complicated diffraction process, for which the interaction with the detector and boundary conditions should be taken into account [47].

The role of various interactions of the obliquely-incident field with the detector is studied in [40]. For instance, interaction of a paraxial beam with an oblique dielectric interface is described by the Fresnel formulas and produces AM-dependent transverse shifts of the reflected and refracted beams similar to (6.25) [33–43]. This is the so-called Imbert-Fedorov shift [136,137].

To consider the simplest model of the field interaction with an oblique detector, we assume that the field reaching the detector surface is totally absorbed and excluded from the beam. In this situation the field-detector interaction can be understood by means of purely geometric arguments; for the scalar beam model, calculations performed with employment of the beam Wigner function (see section 3.2) show [47] that upon an oblique absorbing detection, the beam centroid (3.24) (i.e., defined in terms of the energy density $w$) is shifted with respect to its position in the free field (without detector) by

$$\Delta x_c = m_{xx} \tan\theta , \quad \Delta y_c = m_{xy} \tan\theta \qquad (6.26)$$

where $m_{xx}$, $m_{xy}$ are the elements of the moment matrix $\mathsf{M}_{12}$ (3.32). Displacements (6.26) are written in the beam coordinates $(x, y)$; in fact, they are projections of the shifts measured by the detector in the $(x', y')$-plane. Due to equation (3.32), the elements of $\mathsf{M}_{12}$ (and, consequently, the shifts (6.26)) are associated with the transverse energy flows. In this manner, the 'in-plane' shift $\Delta x_c$ is caused by the radial energy flow (beam divergence) due to which the beam size changes non-uniformly. Indeed, as it is seen in figure 15, before reaching the detector surface $z' = 0$, the 'upper' side of the beam traverses additional distance $\Delta z$ in comparison to the 'lower' side. Hence the upper part of the beam diverges stronger and produces the shift of the centroid along the $x$-axis. In turn, the transverse shift $\Delta y_c$ owes its origin to the azimuthal energy circulation. Indeed, due to the azimuthal component of the internal energy flow, the 'upper' part of the beam experiences an additional lateral motion in the direction of the flow, while the opposite motion of the 'lower' part is blocked by the obstacle (detector plane). For the whole beam this leads to an effective displacement of the beam centroid in the $y$-direction (arrow $C$ in figure 15). Similar effects take place in other situations where "different sides" of a beam traverse different distances, which is typical for boundary refraction or reflection, as it has been mentioned above, and also occur in schemes of grating diffraction [138–140]

The off-diagonal elements of matrix (3.32) quantify the transverse energy transfer related to the OAM (3.33). For a scalar circular vortex beam, e.g., of the type (4.2), whose OAM equals $\mathbf{J}_O = \mathbf{e}_z lW/\omega$ and $m_{xy} = l/2k$, the second equation (6.26) yields

$$\Delta y_c = \frac{l}{2k}\tan\theta \, . \qquad (6.27)$$

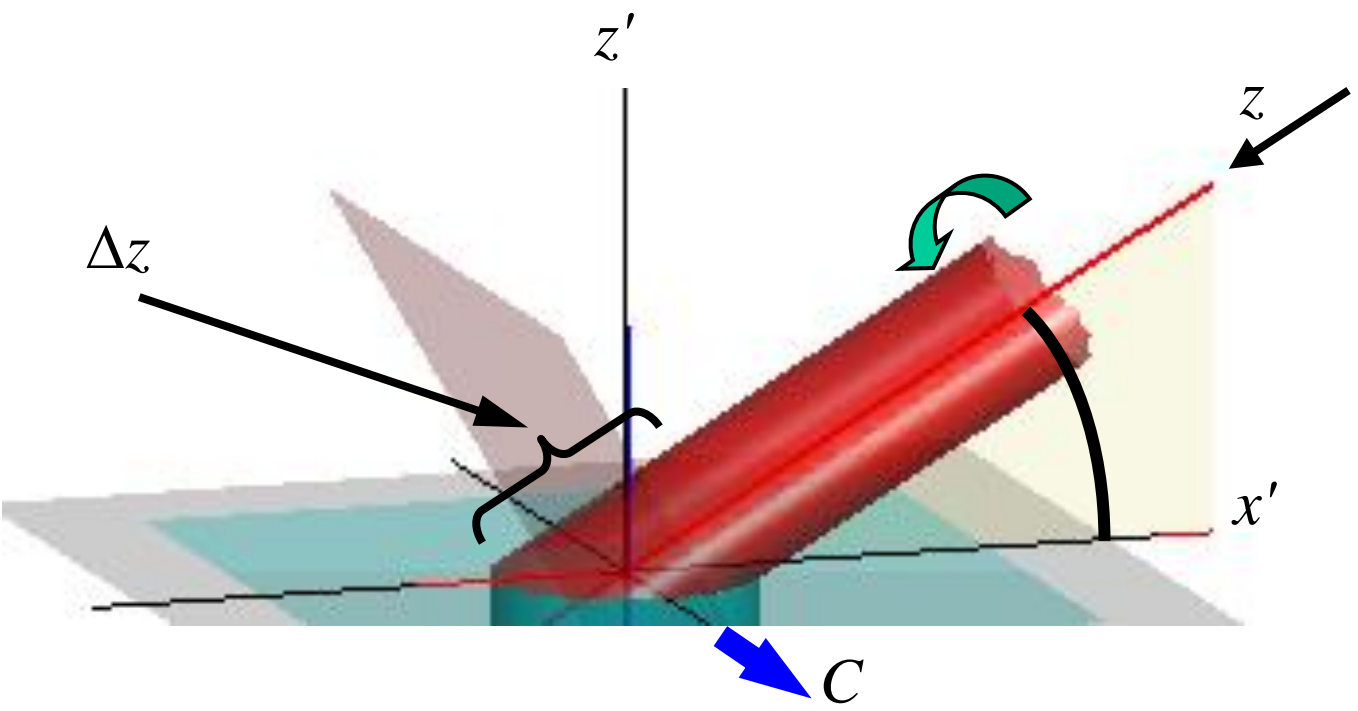

**Figure 15**. Transverse shift of the beam impinging an oblique detector, which is caused by the circular energy flow (see explanations in text).

This result turns out to be quite similar to (6.25), despite the above-mentioned differences in the definitions of the centroid and the boundary conditions caused by the detector. Apparently this coincidence reflects the universal geometric nature of the Hall effects and AM conversion phenomena. Notice also the similarity of the results (6.25) and (6.27) caused, respectively, by the SAM and OAM of the beam. This reveals the dynamical equivalence of the intrinsic SAM and OAM producing spin-Hall [24,33-40,44,46,56,60–63] and orbital-Hall [24,41–43,45,47,58,69] effects in a rich variety of optical systems.

In agreement with the physical meaning of the elements of the moment matrix $\mathsf{M}_{12}$ (section 3.2), the oblique-section techniques offer additional possibilities of studying the transverse energy flows in optical beams [47], including the direct methods for the OAM measurement [140]. They can be suitable in the far infrared and millimetre wave regions where the expected shifts (6.26) are more favourable for measurement than in the visible-light situations.

## 7. Manifestations and visualization of the energy flows

Miscellaneous internal flow patterns considered in the above sections provide a suitable and physically meaningful characterization of optical fields and their spatial structure. However, up to this point, we discussed the internal flows as certain theoretical entities inherent in electromagnetic fields; the important questions of their manifestations and immediate observation were almost never touched upon. This issue is especially important as it is connected to the problem of measurement and experimental investigation of the internal energy flows as well as to their possible applications.

### *7.1. Probing particles*

The most direct manifestations of the internal energy flows are based on their mechanical meaning as the local density of the field momentum (see (2.5)). As a result, if there exists a small particle able to absorb, reflect or scatter the optical waves, in each of these processes the particle's momentum changes simultaneously [4], and this can be detected via the particle motion. This effect is intensively investigated in connection to micromanipulation problems [29,53,141–148]. There are many well developed schemes differing by technical details (sizes and materials of the probing particles, manners they are trapped, suspended in a liquid and exposed to the driving optical field, means for detecting the particle's motion, etc.) but here we only accentuate some principal features of the probing particle approach in application to the optical field diagnostics.

In contrast to early experiments demonstrating the optical AM [1–4,141,142], where probing particles were fixed near the driving beam axis and the field's mechanical action resulted in the particle spinning motion, the optical flows can be seen via the particle trajectory within the beam 'body'. This means that a probing particle is not fixed at the beam axis or other point but should be set free to perform a 3D motion; in most cases where the longitudinal flow is of no interest, the particle is allowed to move within the beam cross section. The particles' orbiting rather than spinning served to distinguish the SAM and OAM contributions [29,53,144]. However, attempts to extract distinct quantitative conclusions from these observations encounter essential difficulties.

7.1.1. The first reason is that the force, with which an electromagnetic field acts on a particle, is not always directly connected with the local field momentum. Even in assumption that the probing particle does not distort the tested field, its action would be quite different, regarding the particle's physical nature [17]: in all cases this action is related with the field momentum but the direct proportionality occurs only for conducting electrically neutral particles. More detailed calculations based on classical [22,146–148] and quantum [1] ideas show that the optical field produces the volume force consisting of the 'dissipative' and 'dipole' parts, both essentially depending on the optical frequency. The 'dipole force' is an odd function of the frequency detuning from the atomic oscillator resonance and depends on the spatial and polarization inhomogeneity of the beam; besides, the gradient force appears responsible for attracting or repelling the particles by regions of high field intensity. The dissipative force depends on the medium absorption coefficient $\alpha$ and is distributed with the volume density [22]

$$\mathbf{F} = c\alpha\mathbf{p} . \tag{7.1}$$

This example, as well as other known analyses [17,146], confirms that under certain conditions the field-induced force can be proportional to the local momentum density. We may estimate the probing particle velocity $\mathbf{v}$, induced by the force (7.1), supposing the particle to be a sphere with radius $a$ suspended within a liquid with viscosity $\eta$. Then the condition of equality between the ponderomotive force (7.1) and the retarding Stokes force, $\mathbf{F}_r = 6\pi\eta a\mathbf{v}$, gives

$$\mathbf{v} = \frac{2}{9}\alpha\frac{a^2}{\eta}c\mathbf{p} . \tag{7.2}$$

Due to (4.8) and (4.9) one can estimate the beam transverse momentum as $|\mathbf{p}_\perp| \sim I/(\omega c b_0)$ and the beam local intensity as $I \sim \Phi/b_0^2$, which immediately results in

$$|\mathbf{v}_\perp| \sim \frac{2}{9}\alpha\frac{a^2}{\eta}\frac{\Phi}{\omega b_0^3} . \tag{7.3}$$

For example, in a focused beam with $b_0$ = 10 μm and wavelength 0.63 μm, a particle with $a$ = 1 μm suspended in water ($\eta \approx 10^{-3}$ kg·m$^{-1}$·s$^{-1}$) will obtain the transverse velocity of the order of magnitude $0.1\alpha\Phi$ μm/s where $\alpha$ is measured in mm$^{-1}$ and $\Phi$ in Watts.

7.1.2. Another important issue is that any object placed in the field disturbs it, and thus any observable mechanical action characterizes properties of the field distorted by the probing particle, rather than the 'original' one. Conditions under which the observable mechanical action can represent the action of the non-perturbed field deserve special consideration [18,146–149]. In the geometric-optics limit the situation is, roughly, understandable: if the absorbing particle 'withdraws the field' from the region of geometric shadow, it experiences the mechanical action proportional to the absorbed momentum. Therefore, in such idealized conditions the particle 'feels' exactly the field momentum which had been concentrated in the volume occupied by the particle before the particle was placed there.

In other cases, as a general model of the field distortion, we can accept that the particle strongly disturbs the field in the shadow region, plus certain transition zone of the near-wavelength size [18]. For particles of near-wavelength and, especially, sub-wavelength sizes, this leads to a conclusion that the field-particle interaction can be much stronger as well as much weaker than it follows from the naïve geometric considerations (see figure 16 for illustration [149]). As a matter of fact, for sub-wavelength particles the force acting on a particle generally has little in common with the local momentum of the unperturbed field, even if the physical conditions, at which the field mechanical action is exactly proportional to the Poynting vector (see paragraph 7.1.1 above), are realized.

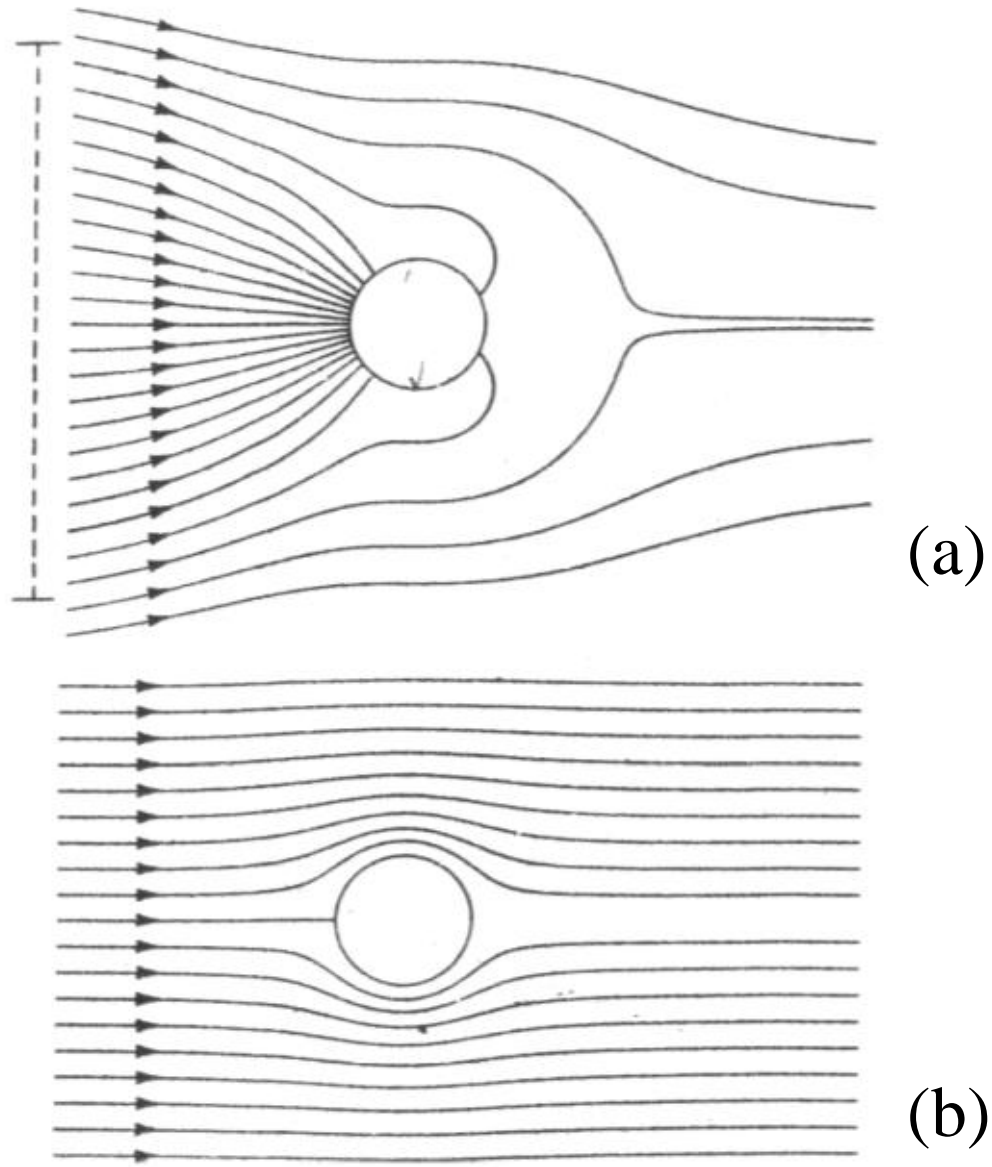

**Figure 16**. Poynting vector lines near the aluminium sphere of 20 nm in diameter for different radiation wavelengths [149]: (a) $\lambda$ = 140 nm, (b) $\lambda$ = 220 nm. In panel (a) the size of the particle-induced disturbance is shown by the dashed line.

In this view, one may conclude that probing particles can serve to studying the internal flows only in large enough spatial scale ($\gg \lambda$). So, the physical observability of the sub-wavelength structures in the Poynting vector distributions appears to be questionable (see footnote 1 in section 4.3). Such structures can be detected only with sub-wavelength probing particles but any such particle distorts the field in the region of at least wavelength size [18,149]: in any thinkable experiment the field momentum inhomogeneities with size $< \lambda$ seem to be effectively averaged and cannot be resolved by means of the particle motion.

A special reservation should be made in respect to the spin flow observations. Even in the geometric-optics conditions, any particle perturbs the field intensity stronger than the phase. This weakly affects the OFD field “before” the particle (in the half-space the light waves come from) because due to (3.11) and (3.44) it is determined mainly by the phase distributions of the field polarization components. On the contrary, the SFD pattern (3.36) depends on the intensity distribution, and is, generally, strongly deformed by the particle presence. For example, the field intensity gradually grows along the outer normal near the surface of an absorbing particle. In such cases the SFD lines, oriented along the contours of constant intensity (see figure 1), will flow around the particle and, seemingly, exert no force on it. However recent model calculations for simple field configurations [148], employing the Mie theory [149], seem to confirm the mechanical action of the SFD, though with rather peculiar behaviour depending on the particle size and optical properties.

7.1.3. There is one more difficulty in quantitative interpretation of the ‘probing particle’ experiments that is not as principal as the above ones but is important from the technical point of view. The matter is that, besides the electromagnetic field, the probing particle is subject to many accompanying factors and actions of different nature [141,146]: viscosity of the suspending liquid, the cell-wall friction, gradient forces due to the field inhomogeneity, ‘propeller’ forces stipulated by the particle shape, etc., whose combined action is beyond all calculation. In any case, the expected mechanical action is rather weak; the above analysis and equations (7.2) and (7.3) clearly witness for the effect detectability as well as for practical difficulty to retrieve the detailed flow patterns in this way. These circumstances prevent from unambiguous quantitative interpretation of the experiments even if all the above conditions necessary for the one-to-one correspondence between the mechanical action of the field and its momentum density are satisfied.

To summarize this sub-section, we ought to emphasize that the mechanical action on probing particles provides unique possibility to inspect the internal flow pattern but its real potential is limited to relatively smooth fields (in any case, the spatial resolution is larger than the wavelength), and even in such cases, only the qualitative representation of the flow distribution is available. However, new approaches to the field diagnostics via scattering by a single atom or molecule [150] may serve to overcome the above limitations.

*7.2. Free-space transformation of the beam profile.*

As was shown in section 3.4, the transverse energy flows in a paraxial beam immediately manifest themselves in transformations of the visual beam profile upon its free propagation. However, this process is regulated by relation (3.46) which gives access only to the 2D divergence of the transverse flow field $\mathbf{p}_{\perp}$. As a result, the divergenceless part of the flow remains ‘hidden’. This is a severe restriction because due to (2.11) the solenoidality is an inherent property of the optical flows; in particular, this approach can give no information on the SFD distribution.

Even in application to the orbital flow this approach requires to restore a vector field from its divergence, which is an ambiguous and potentially misleading operation. This can be easily seen on the behaviour of so called ‘spiral beams’ [151–154] – a class of paraxial beams with self-similar rotatory propagation whose intensity profile retains its shape, except enlargement due to the self-diffraction and rotation near the propagation axis. It is tempting to associate the visual beam profile rotation with the azimuthal component of the orbital flow whose integral characteristic is the OAM (see section 3.4). However, it has been noticed that the visual profile rotation can occur in beams with zero OAM [155,156] and even rotation handedness opposite to the OAM handedness may occur [155]. Figure 17 illustrates the intensity profile evolution of two paraxial beams formed by simple superpositions of the standard LG modes (4.2). Both beams possess the same OAM per unit power but their intensity profiles rotate oppositely.

The physical reasons for such behaviour can be understood from figure 18 displaying maps of the transverse energy flows in these beams. Indeed, in both cases the overall circulatory flow is directed counter-clockwise, which seems to testify that the whole pattern must have been ‘transported’ also counter-clockwise. But in figure 18b, the energy current lines converge when approaching the intensity maximum and diverge on coming out of it. That is why the energy concentrates in the ‘rear’ and dissipates near the ‘front’ of the bright spot. This process is superimposed on the ‘normal’ energy transport, and the summary effect, determined by the competition, is the ‘backward’ rotation. In figure 18a, the energy flow convergence and divergence act in agreement with the ‘normal’ energy transport and increase the rotation velocity.

The described picture of the transverse energy flows enables to explain apparent discrepancies between the visual beam rotation and the predominant direction of the transverse energy circulation. Let an observer watch, for example, the beam transformation from pattern of figure 17a to that of figure 17b. At first glance, it looks as if the light energy moves from a certain initial point, say, A to its current position at point A' along the arc AA'. Such way of reasoning implicitly supposes that the beam “rotates” like a rigid body, which, in view of the above remarks and section 4.1, is generally incorrect. On the contrary, the visible transformation of A into A' can be realized if some portion of the beam energy moves from A towards the beam axis O, and the equivalent portion goes from the central area to A', approximately as is shown by white arrows in figure 17a. Comparison of figures 17a, 17b and figure 18a shows that real picture of the energy transfer is close to this schematic which, of course, goes without any azimuthal flow.

Quantitatively, transformation of the intensity pattern of a propagating beam is described by relation (3.46) that can be rewritten in the form

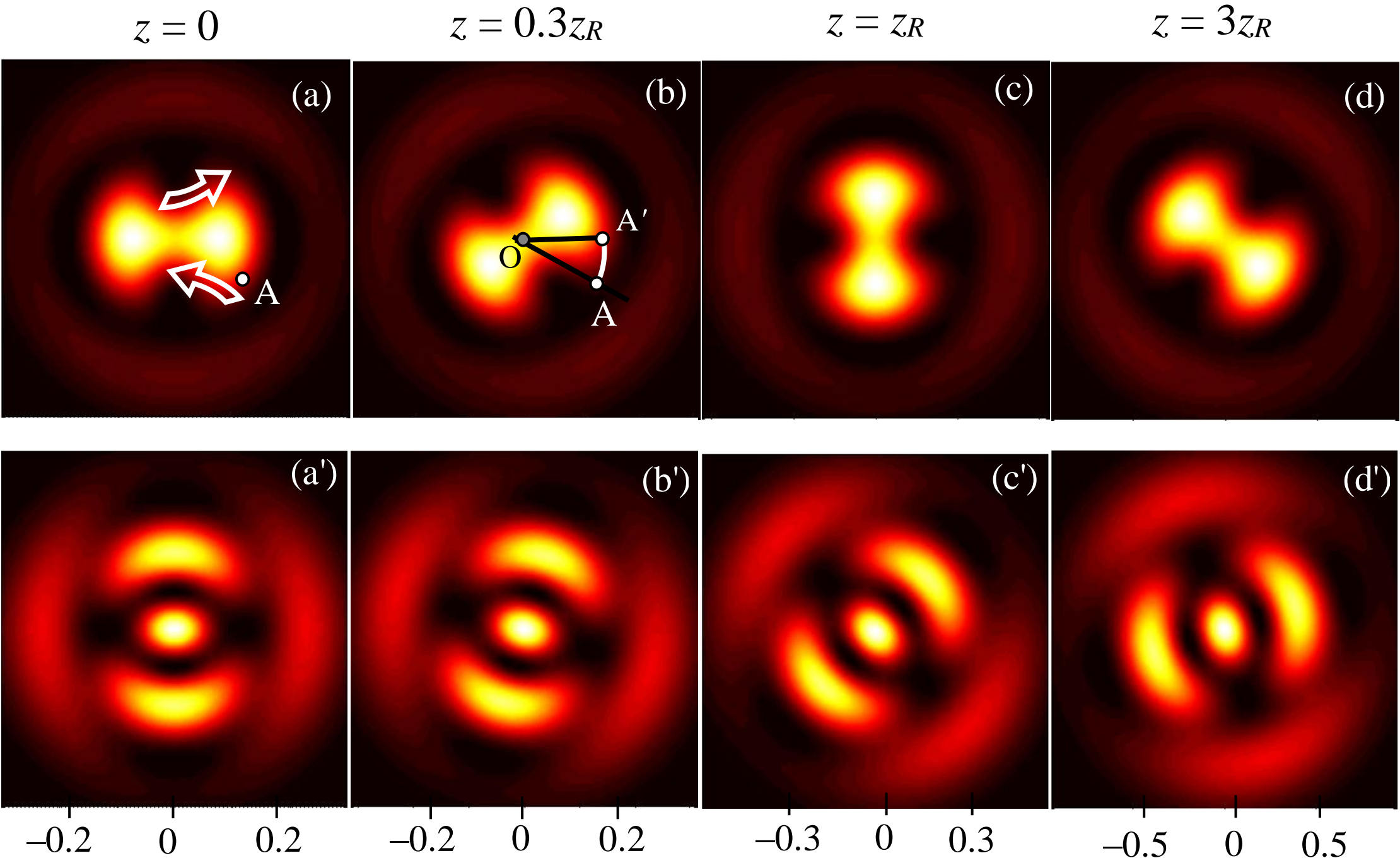

**Figure 17**. Evolution of the transverse intensity pattern of model spiral beams (superpositions of LG modes with equal powers, see section 4.1) viewed against the beam propagation: (a)–(d) superposition $u_{12} + u_{00}$; (a')–(d') superposition $u_{02} + u_{20}$. Distances from the initial (waist) plane are indicated; the beam broadening can be traced by the scale labels in the lower panels (in centimeters). Details in panels (a), (b) are explained in the text; the beam parameters accepted in calculations are: $b_0$ = 1 mm, $k = 10^5$ cm$^{-1}$ (He-Ne laser).

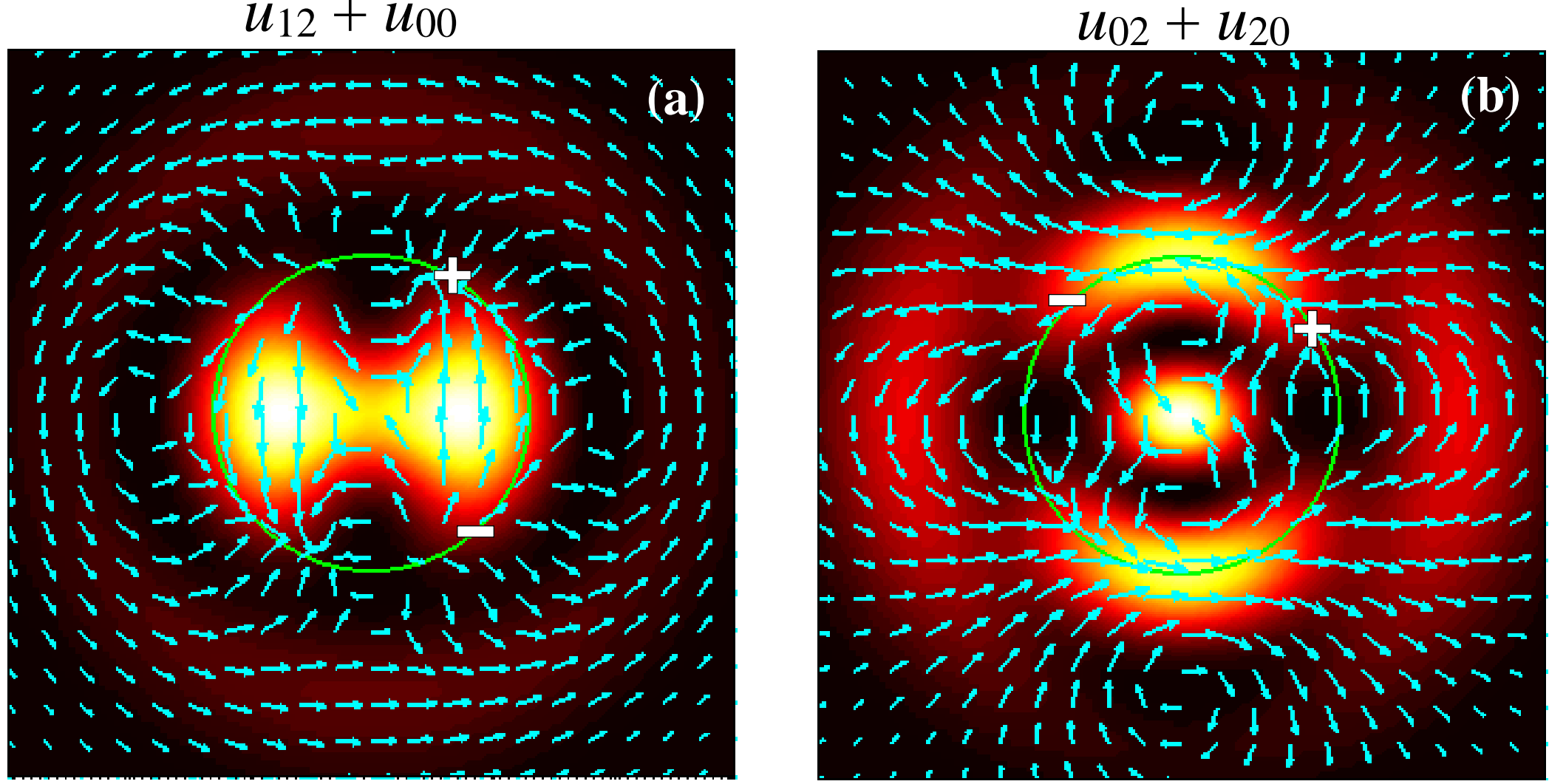

**Figure 18**. Spot patterns of spiral beams of figure 17 with OFD maps, signs "+" ("–") denote regions where flow lines converge (diverge). Light circumferences are examples of closed contours surrounding the beam axis (see reference [155]).

$$c\frac{\partial w}{\partial z} = -\mathrm{div}\left(w\mathbf{v}_{\perp}\right) \tag{7.4}$$

where the 'transverse energy flow velocity'

$$\mathbf{v}_{\perp} = c^2 \frac{\mathbf{p}_{\perp}}{w} \tag{7.5}$$

is introduced. Equations (7.4), (7.5) form a ground for the 'hydrodynamic' approach to the beam evolution that appears especially helpful in complex situations of stochastic wave propagation [157–159]. Note that by introducing the electromagnetic mass with the density $m_e = w/c^2$ we can formulate the law (7.4) of the beam transformation exactly in terms of the fluid mechanics; however, this analogy is not complete because, except the continuity equation (7.4), the electromagnetic 'fluid' also obeys the Maxwell equations rather than mechanical equations of the fluid motion.

*7.3. The beam constraint, symmetry breakdown and the transverse energy flow*

Material of the previous section witness that comparison of the paraxial beam profiles in different cross sections can give information on the internal energy flows. At the same time, in case of freely propagating beams this information is rather ambiguous and, generally, provides no definite conclusions about the transverse momentum distribution. To make the results more specialized, one can try to consider the beam propagation in conditions of transverse constraint or deliberate deformation of the beam profile. When a beam meets a properly shaped obstacle, its profile is distorted but the transverse momentum distribution in the half-space where the beam comes from remains more or less the same. In the course of further propagation (after the obstacle) the transverse energy redistribution goes in accord with the initial velocity (7.5) so that it can be visualized. This effect is often coupled with the symmetry breakdown of the initial circular beam due to which the transverse transport of energy comes to light and can be suitably observed [77,79].

In the extreme case, the beam transverse section is fractionated into a number of 'beamlets' that propagate in directions determined by the local transverse momentum density. This concept is employed in regular approaches to the wavefront sensing such as the Hartmann method [160]; in essence, they measure the transverse orbital momentum [158] whence the phase profile is reconstructed via (3.44) or (3.45). However, this approach is only applicable to rather wide and smooth beams that occur, e.g., in astronomy or in the optical testing, including the vision correction. For more complicated beams the same ideas can be implemented in less immediate but still useful ways.

An impressive example of the transverse flow visualization is realized in the scheme of partial screening the circular $LG_{01}$ beam [161] (figure 19). In the experiment, the beam was linearly polarized so the incident field contained only the orbital transverse flow determined by second relation (4.9). After the obstacle, the beam energy which is not absorbed by the screen (in the bright region above the screen edge in figure 19) continues to move in agreement with the 'initial' momentum obeying (7.4) and (7.5). The azimuthal flow existing in the non-perturbed LG beam causes that the beam energy visually 'moves' into the shadow region along the spiral trajectory.

Quite similar manifestations of the azimuthal flows occur in other situations when a certain part of the beam cross section is artificially isolated by spatial obstacles. In particular, the well known patterns of diffraction by a slit [162,163] and by a non-transparent stripe [164,165] in beams with azimuthal transverse orbital flow are deformed (figure 20): The interference fringes bend in agreement with the local transverse momentum (or, which is the same, with the local velocity of the transverse energy transfer (7.5)). Of course, all these processes can be described in the diffraction language which is used in the original works [161–165] but employment of the transverse flows makes the behaviour physically clear and fairly spectacular.

In the case of a 'soft' slit-like diaphragm with Gaussian transparency profile

$$T(\mathbf{r}) = \exp\left(-y^2/d^2\right), \tag{7.6}$$

the interference fringes do not emerge and the azimuthal transfer of the beam energy is clearly observable via consecutive spot patterns of the propagating beam (figure 21 [77]). The visible rotational behaviour of the beam profile is regulated by combination of the azimuthal and radial (second term of (3.45)) energy flows.

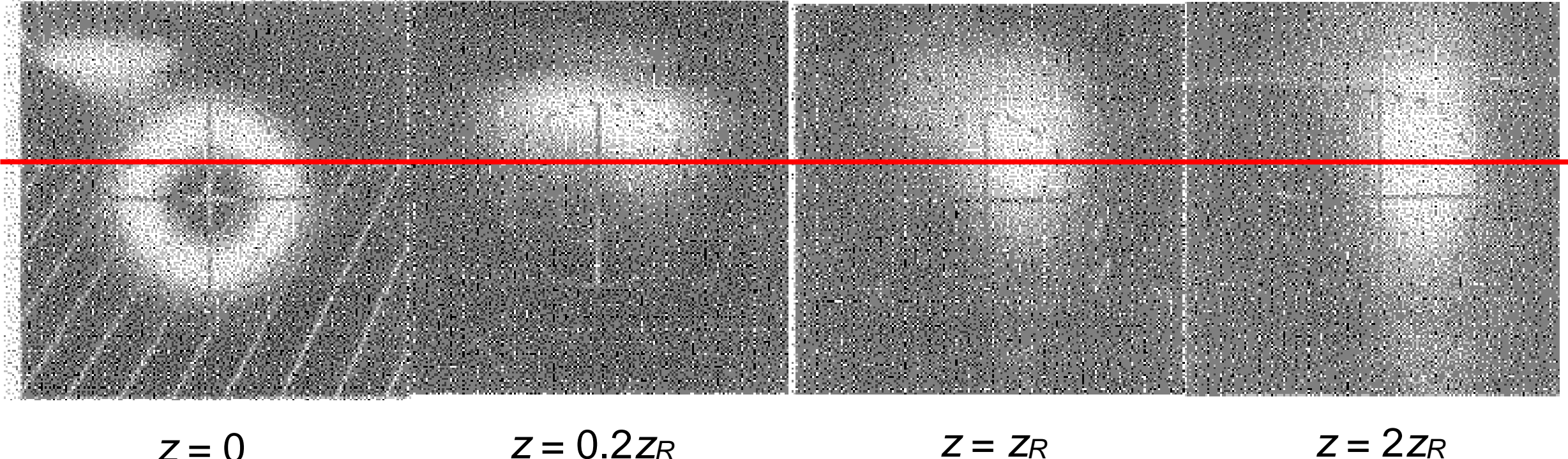

**Figure 19**. Lateral shift due to the circular energy flow in the partially screened $LG_{01}$ beam. Horizontal line: projection of the screen edge, dashed lines: contours of the non-perturbed LG beam; propagation distances are shown in units of the Rayleigh range (4.4).

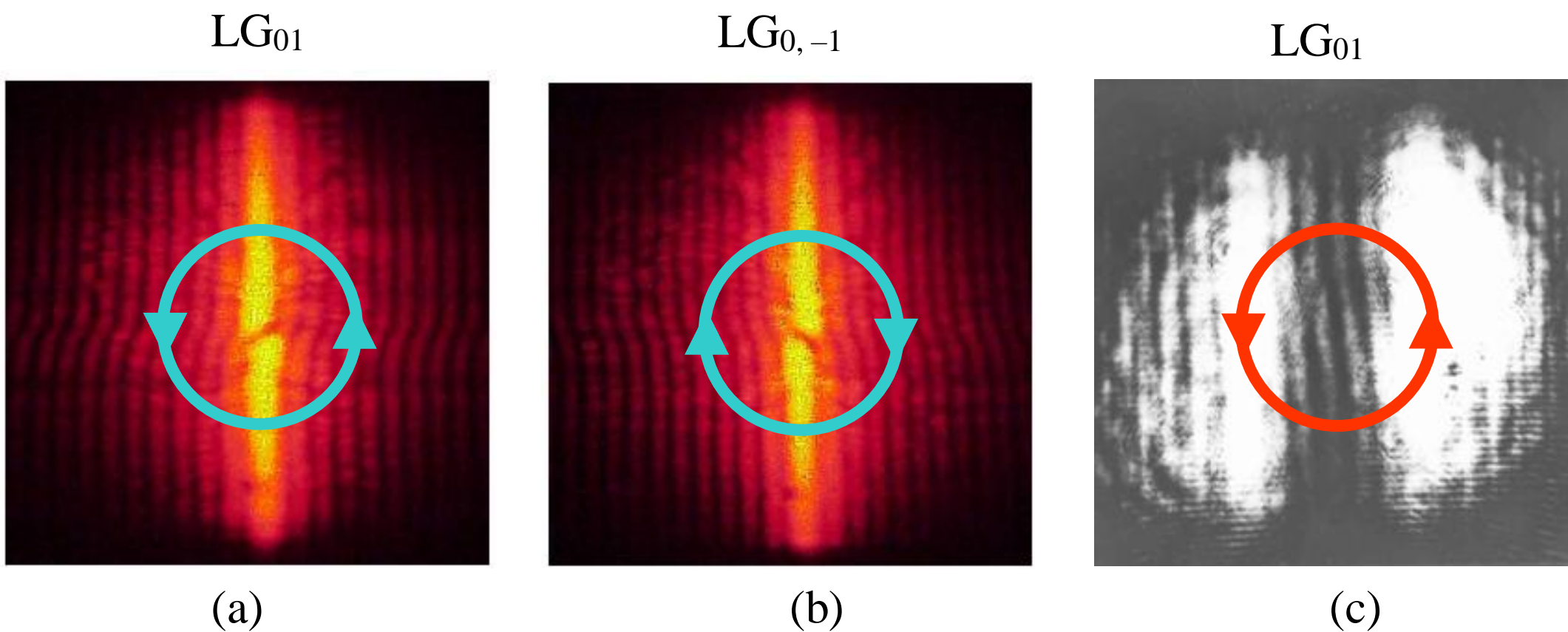

**Figure 20**. Visualization of the azimuthal energy flow by bending the interference fringes: (a) and (b) the slit diffraction [162]; (c) non-transparent strip diffraction [164]. Oriented circles show the circulatory flow direction.

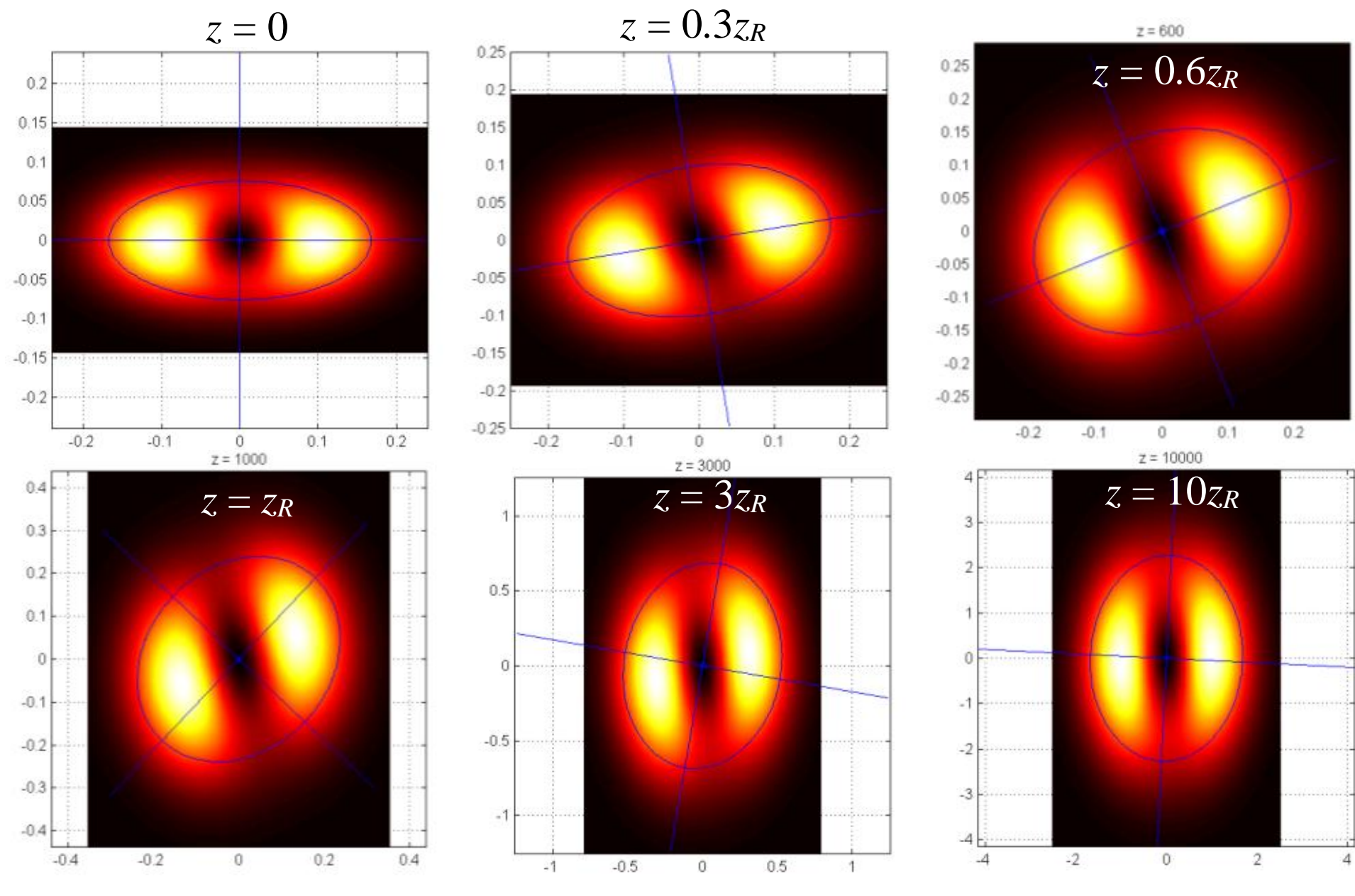

**Figure 21**. The beam pattern variation during the propagation of the $LG_{01}$ mode of equation (4.2) with the initial beam size $b_0$ = 0.1 cm, after passing the Gaussian diaphragm (7.6) with $d$ = 0.07 cm. The diaphragm is situated at the beam waist, propagation distances are indicated in units of the Rayleigh range (4.4); the images' sizes are normalized by the current transverse scale of the beam.

The beam transformations described in the above paragraph can be considered as realizations of symmetry breakdown of circular beams. Other examples of the same principle (LG beam transformations performed by an astigmatic lens [126], asymmetric telescopic transformations [45,127,128], or optical vortex generation with a misaligned 'fork' hologram [166]), confirm the general rule that visible evolution of the transformed beam spot carries the information on the internal energy flows before the symmetry breakdown. However, in most cases this relation is rather complicated and can only be used for qualitative indication of certain features of the energy flows, e.g., for detection of the azimuthal energy transfer and its predominant direction [77,161,164,165].

Choosing of an oblique section of the beam for the analysis (section 6.2) can be considered as a sort of its symmetry breakdown and, as well, can serve to extract the information of the internal transverse flows [47]. Equations (6.26) immediately give access to the elements of the moment matrix which, due to (3.32), are related to the transverse momentum distribution. This approach seems to be productive but needs further development. In practice, it could be expedient to employ several differently oriented oblique sections of the beam and then combinations of results obtained for each case open a way to more complete characterization of the transverse energy flows. The centroid measurement is only the simplest manipulation with the beam profile and, probably, more comprehensive comparative study of the beam profiles in the transverse and oblique sections will enable to 'spy' the internal energy flows within the beam with much more details than the second irradiance moments of (3.32) cover. Of course, this requires rather high measurement accuracy which can easier be achieved for far-infrared or millimetre wavelength regions.

Another line of generalization of the oblique section approach can be associated with the use of non-planar secant surfaces instead of oblique planes. In this way, the beam profiles in various curvilinear sections can be considered, which supplies additional instruments for characterizing the transverse energy flows in certain selected regions of the beam cross section [47].

## 8. Conclusion

Characteristics of the internal energy flows (or, equivalently, electromagnetic momentum density) in light beams, as well as their partial contributions belonging to separate polarization components, constitute a physically meaningful and application-oriented set of the beam parameters. Additionally, they provide a deeper penetration into the 'intimate' processes associated with the light beam propagation and transformations. Although at present time the main properties, descriptive potential and application conditions of the internal flows are not completely clear, the known facts and concepts form a consistent interrelated system that can be summarized as follows.

1. The total energy flow of a light beam is described by the Poynting vector field. It can be divided into three 'structural' summands associated to the three degrees of freedom corresponding to the extrinsic AM, intrinsic orbital AM and intrinsic spin AM of the beam. The extrinsic AM owes to transverse displacement of the light energy of the beam 'as a whole'; in turn, the intrinsic energy redistribution consists of two structural parts with different features. The spin flow is related to the vector nature of light (polarization) and vanishes if the polarization is linear; it emerges from the inhomogeneous distribution of the third Stokes parameter (degree of circular polarization) and possesses zero divergence. The orbital flow originates from the spatially inhomogeneous distributions of the amplitude and phase of separate polarization components. It exists for scalar as well as vector optical fields and is responsible for transformation of the visible beam profile in the course of free propagation. Both intrinsic structural components of the energy flow can be attributed to the peculiar instantaneous 'motion' of the electromagnetic field that occurs with optical frequency and possesses no direct mechanical meaning. For the spin flow, this motion is obviously the instantaneous rotation of the field vectors; for the orbital one, it is the 'running' component in the pattern of instantaneous oscillation of the spatially inhomogeneous electromagnetic field.

2. The generic properties of the internal energy flows are illustrated by numerous analytic and numerical calculations related to the standard beam models (Gaussian, Laguerre-Gaussian, Bessel, flat-top beams) and to general models of spatially inhomogeneous beams, including the stochastic ones. The most impressive feature is helical and even circular flow distributions that occur in the orbital flow patterns near lines of zero intensity. Such structures were intensively studied during the past years in relation to the optical vortices. In essence, analysis of the flow patterns in the scalar beams shows that the energy flow in an optical vortex can be interpreted as a mechanical vortex motion of a fluent “body” whose density is determined by the mass equivalent of the beam energy. In vector beams, the spin flow circulation takes place near extrema of the third Stokes parameter distribution. The spin and orbital flows can mutually support or compensate each other, up to vanishing of the transverse momentum at certain lines and surfaces within the beam field.

3. Singularities of the internal flows occur where the Poynting vector or its transverse component vanishes. In the 3D patterns, the singular points can form 0D (isolated points), 1D (lines) and 2D (surfaces) manifolds. In the most interesting case of the transverse flow fields, the classification of singular points is common for any 2D vector fields and includes nodes (sinks and sources), saddle points, vortices (circulation points) and attractive or repelling focuses (spiral points). In scalar beams, the flow singularities generally coincide with corresponding phase singularities. For vector beams, the singularities of the partial flows, belonging to separate polarization components, in many cases can be associated with the usual polarization singularities (C-points and L-contours). However, positions and classes of singularities of the total transverse flow generally cannot be related to certain polarization singularities and only some correlations are being sought in the current research.

4. The different structural parts of the energy flow can be mutually converted in the processes of the beam transformation. This can be treated as the momentum and energy redistribution between the different degrees of freedom of the light beam. Within the frame of this review restricted to the free-space effects, the spin-to-orbital flow conversion upon the beam focusing (defocusing) and upon the beam symmetry breakdown (especially, due to observing an oblique section of the beam) is discussed. In the first situation, the circular polarization of the initial collimated beam contributes to the helical wavefront in the longitudinal component of the focused beam with corresponding vortex orbital flow. In the second one, the transverse energy flows, ‘balanced’ in the normal cross section, produce a ‘disbalance’ in the energy distribution over the oblique section, depending on the conditions for the light propagation between the normal and oblique sections. Both effects are examples of the spin-orbit interaction and can serve to manifest the internal flows.

5. However, the problem of measurement or, wider, visualization of the internal flows in light beams remains insistent. The known methods for the flow pattern observation serve to its qualitative detection rather than to quantitative evaluation. In principle, the energy flows can be determined via measured distributions of amplitudes and relative phases of separate polarization components [113] (e.g,, by means of the spatial-resolution Stokes-polarimetry [167,168]) and subsequent application of the basic formulas from sections 2 and 3. But this is an indirect approach. More promising way is based on the mechanical action of the optical field upon suspended microparticles that absorb or reflect some part of the beam momentum. Also, the internal flows can be determined via the intensity profile transformation during the beam propagation in the free space as well as in conditions of spatial constraint (special transparencies, masks, etc.) or purposeful transverse deformation. One of such means (the Hartmann method and its variations) works quite reliably but only for beams with very smooth inhomogeneity and in large spatial scales; other methods permit to obtain only qualitative results, which in many cases are difficult to interpret.

Essential difficulties arise in trying to immediately ‘peering’ the sub-wavelength structures in the energy flow patterns. To our opinion, this fact is a special aspect of the more general fundamental problem relating the properties of the energy flows in the sub-wavelength scale. There is no common opinion on the interpretation of sub-wavelength inhomogeneities in the time-averaged Poynting vector distribution [17]; in this view, the early attempt of introducing the space-averaged Poynting vector to represent the internal energy flow [15,16] seems to deserve additional attention.

The specific properties of the spin flow component (see sections 7.1, 7.2) can make an impression that it 'escapes' from direct observations, so the question of its observability needs the special discussion. We hope it can be resolved by experiments aimed to find the orbital motion of particles localized at the contours of a high gradient of the third Stokes parameter $s_3(\mathbf{r})$ within beams with a plane wavefront and, therefore, with no transverse orbital flow (e.g., 'hybridly polarized vector beams' of [169]).

6. To finalize the present consideration of internal flows in light fields we would like to emphasize once again that treating a light field as a complex of the energy currents provides a clear and physically meaningful representation of important properties associated with the 'fine structure' of the beam. Many subtle effects, in particular, those related to the spin-orbit and orbit-orbit interactions that look, at first glance, 'counter-intuitive' [46], become quite understandable and even expectable when the internal energy flows are taken into account. The spin and orbital momentum densities, their partial contributions belonging to the separate polarization components, energy and momentum distributions with their auxiliary characteristics (the irradiance moments and the centroid trajectory) are theoretically irreproachable parameters reflecting the most fundamental dynamical and geometrical aspects of the optical fields. Simultaneously, they proved to be valuable heuristic instruments for studying the light beam transformations, especially suitable in processes involving the light angular momentum and interactions between different rotational degrees of freedom of light. They also supply immediate and meaningful characterization of the light beams in terms appropriate for many applications, from the information transfer up to micromanipulation. It is worth noticing that the energy density and energy flow have useful counterparts – the chirality density (spin energy) and chirality flow (SAM density) [170–173]. They play an important role in optical interaction with chiral particles [172] and, together with other quantities, form Poincaré invariants of an electromagnetic field [170].

## Acknowledgements

This work was supported, in part, by the Ministry of Science and Education of Ukraine (project No 457/09), European Commission (Marie Curie Action), and Science Foundation Ireland (Grant No. 07/IN.1/I906). The authors are grateful to Andrea Aiello, Miguel Alonso, and Michael Berry for fruitful discussions and correspondence.